\documentclass[letterpaper,twocolumn,10pt]{article}
\usepackage{usenix-2020-09}

\usepackage{tikz}
\usepackage{amssymb}
\usepackage[font=small]{caption}
\usepackage{amsmath}
\usepackage{xspace}
\usepackage{enumitem}
\usepackage{wrapfig}
\usepackage{subfig}
\usepackage{float}
\usepackage{libertine}
\usepackage{listings}
\usepackage[ruled, linesnumbered, vlined]{algorithm2e}
\usepackage{color}
\usepackage{booktabs}

\newcommand{\SmartGen}{\textsf{\textsc{SmartGen}}\xspace}

\newcommand{\ie}{{\em i.e.},\xspace}
\newcommand{\eg}{{\em e.g.},\xspace}

\usepackage{titlesec}
\titlespacing\section{0pt}{4pt plus 4pt minus 3pt}{4pt plus 2pt minus 2pt}
\titlespacing\subsection{0pt}{4pt plus 3pt minus 2pt}{4pt plus 2pt minus 2pt}
\titlespacing\subsubsection{0pt}{4pt plus 3pt minus 2pt}{4pt plus 2pt minus 2pt}
\usepackage{url}
\usepackage{xurl}

\usepackage[affil-it]{authblk}
\makeatletter
\renewcommand\AB@affilsepx{\quad\quad \protect\Affilfont}
\def\thanks#1{\protected@xdef\@thanks{\@thanks
        \protect\footnotetext{#1}}}
\makeatother

\begin{document}
%-------------------------------------------------------------------------------
%don't want date printed
\date{}

\title{\Large \bf \SmartGen: Seamless Disaggregated LLM Inference with\\ Selective KV Cache Transfer}
% \title{\Large \bf \SmartGen: Efficient KV Cache Transfer for Self-Hosted, Disaggregated LLM Inference Systems}
% \title{\Large \bf \SmartGen: Seamless Disaggregated Inference for Self-Hosted LLMs with Selective Key-Value Cache Transfer}
% \title{\Large \bf Offering Seamless Disaggregated Inference for Self-Hosted LLMs with \SmartGen}
% \title{\Large \bf \SmartGen: Seamless Disaggregated Inference for Self-Hosted LLMs}
% \title{\Large \bf \SmartGen: Seamless Prefill-to-Decoding in Disaggregated Inference for Self-Hosted LLMs on the Cloud}
% \title{\Large \bf \SmartGen: Seamless Stage Transition in Disaggregated Inference \\ for Self-Hosted LLMs on the Cloud}

%for single author (just remove % characters)
% \author{\rm Submission ID: 381}
\author[1]{Xuchuan Luo}
\author[2]{Jiacheng Shen}
\author[1]{Xin Wang}
\author[1]{Yangfan Zhou}

\affil[1]{College of Computer Science and Artificial Intelligence, Fudan University\authorcr}
\affil[2]{Duke Kunshan University}

\pagestyle{plain}
\maketitle

\begin{abstract}
\noindent
Disaggregating the prefill and decoding stages of large language model (LLM) inference into two separate sets of nodes is widely adopted in today's LLM serving systems. 
However, such an architecture poses significant challenges for self-hosted LLM deployments on rented cloud instances, since transferring enormous key-value (KV) caches between disaggregated nodes can easily saturate the limited inter-node network bandwidth.
% Disaggregating the prefill and decoding stages of large language model (LLM) inference into two separate sets of nodes is widely adopted in today's LLM serving systems.
% However, such an architecture induces significant burdens on the network bandwidth when hosting LLMs on rented cloud instances,
% \ie transferring enormous key-value (KV) caches inevitably saturates the limited network bandwidth, thereby stalling the stage transition and degrading the user experience.
% and hinders the efficiency of the decoding stage.
% 引出本文和挑战
% In this paper, we propose to relieve the burden by selectively transferring essential KV cache entries across the two stages.
In this paper, we propose to mitigate the network bottleneck by selectively transferring essential KV cache entries across the two stages.
There are two challenges to achieve selective KV cache transfer, \ie accurate KV selection during the prefill stage, and efficient KV fetching during the decoding stage.
To address these challenges, we design \textbf{\SmartGen}, a KV cache transfer engine that allows seamless disaggregated LLM inference with three data transfer paths.
Specifically, we leverage 1) a \textit{profile-based proactive transfer} path to identify and push essential KV cache entries to the decoding node during the prefill stage, 2) a \textit{parallel on-demand transfer} path to simultaneously fetch remote and local KV cache entries during the decoding stage, and 3) a \textit{speculative transfer} path to finally deliver all KV caches to the decoding node.
% 一句话实验结果
Experimental results show that \SmartGen reduces time-to-second-token by up to $4.3\times$ compared with the typical full KV cache transfer approach while offering comparable subsequent decoding performance and accuracy.
\end{abstract}
\section{Introduction}
% 先说最新的大模型推理系统采用KVCache-centric的PD分离，带了来xxx好处
\noindent
Large language models (LLMs) have been widely deployed in real-world applications, \eg chatbots~\cite{chatgpt, fast2025mooncake}, programming assistants~\cite{arxiv2021llm-on-code, arxiv2023llama-on-code}, and summarization tools~\cite{acl2023sum, expert2022sum}, thanks to their powerful capabilities in language understanding and generation.
Existing LLM inference systems disaggregate the prefill and decoding stages of inference computation into separate sets of nodes to accommodate their distinct computational characteristics~\cite{osdi2024distserve, fast2025mooncake, isca2024splitwise, nvidia-dynamo, arxiv2024deepseekv3}.
This enables independent and flexible resource allocation for each stage, maximizing the overall throughput.

% 然鹅，decoding-side network bandwidth是disaggregated的瓶颈
% However, the disaggregated scheme hinders individual or enterprise users~\cite{aws-partners} from deploying self-hosted LLM inference systems due to the network bandwidth bottleneck.
However, the disaggregated scheme hinders users~\cite{aws-partners} from deploying self-hosted LLM inference systems due to the network bandwidth bottleneck.
% 原因有二: 1) kv cache线性增长 2) 云服务带宽limited
Specifically, key-value (KV) caches, \ie intermediate attention states, must be transferred from the prefill nodes to the decoding nodes when serving each inference request.
Existing systems~\cite{fast2025mooncake,isca2024splitwise,icml2024dejavu} transfer KV caches in a layer-by-layer manner to overlap data transfer with prefill computation.
However, according to our experiments on Llama~\cite{llama-3} and Qwen~\cite{qwen3} models on Alibaba cloud GPU instances with L20 GPUs and 25 Gbps RDMA network, transferring long-context prompts, \ie 48K tokens, takes $6.5\times$ more time than the prefill computation.
The reason is twofold.
On the one hand, since the size of the KV cache is large and grows proportionally with the sequence length, batch size, and model size, it inevitably saturates the limited network bandwidth~\cite{osdi2024distserve, isca2024splitwise}.
On the other hand, most cloud GPU instances only offer limited network bandwidth~\cite{aws-instances, google-instances, alibaba-cloud-instances, tencent-cloud-instances}, \eg the L4 and L40S instances of AWS~\cite{aws-instances} and A100 instances of Google Cloud~\cite{google-instances} provide only 10–35 Gbps network.
% According to our experiments on Llama~\cite{llama-3} and Qwen~\cite{qwen3} on Alibaba cloud GPU instances with L20 GPUs and 25 Gbps RDMA network, transferring XXXX prompt, \ie 48K tokens takes $6.5\times$ more time than the prefill computation.
% 引出后果，定义问题
% Such a bottleneck hinders the decoding stage, thereby impacting the user experience with increased \textit{time-to-second-token} (TTST) and underutilizing resources on the decoding node.
The KV cache transfer hinders the decoding stage and can account for up to 42.2\% of the job completion time~\cite{sigcomm25hack}, hurting the user experience with increased \textit{time-to-second-token} (TTST).
We refer to this problem as a \textit{stage-transition stall}.

% 现有方案通过prefill来掩盖传输时间，然而这种方法不能从根本上解决问题，因为xxx
% Existing systems~\cite{fast2025mooncake, isca2024splitwise, icml2024dejavu} propose transferring KV caches layer by layer during the prefill stage to overlap data transfer with prefill computation.
% However, our experiments on Llama~\cite{llama-3} and Qwen~\cite{qwen3} show that, with L20 GPUs and 25 Gbps RDMA network, transferring 48K KV cache tokens takes up to $6.5\times$ more time than the prefill computation after adopting various prefill optimizations, \eg prefix caching~\cite{fast2025mooncake, arxiv2025cachewild, nips2024sglang}.
% Nonetheless, as various optimizations are adopted in the prefill stage~\cite{nips2022flashattention, nips2024sglang, fast2025mooncake, fast2025impress, atc2024cachedattention}, network latency can hardly be overlapped by the faster prefill computation.
% the prefill stage often executes too fast to fully hide the KV cache transfer.
% Our experiment on Llama and Qwen models~\cite{llama-3, qwen3} shows that, with L20 GPUs and 25 Gbps RDMA network, KV cache transfer for 48K tokens takes up to $6.5\times$ more time than the prefill computation.

In this paper, we propose to accelerate KV cache transfer by leveraging the inherent importance distribution of KV caches, \ie not all KV entries are required during the decoding stage.
% 现有工作普遍使用 KVCache 稀疏性来加速 LLM 推理
Specifically, existing works have widely exploited the dynamic sparsity in KV caches to accelerate the decoding computation~\cite{osdi2024infinigen, acl2025hata, nips2023h2o, arxiv25retroinfer, deepseek-dsa}.
% 如果能在 prefill-decoding 之间的 KVCache 传输阶段使用 sparsity，那么就能显著减少传输数据量，降低传输时延
If the sparsity can be utilized in the transfer phase, the network bottleneck can be significantly relieved since only a small amount of important KV entries need to be transferred during the prefill stage.
% 然而，要实现这个 idea 需要解决以下几个挑战
However, two challenges must be addressed before this idea becomes a practical KV cache transfer solution.

% 本文提出XXX的做法来解决问题，解释为什么能解决，引出挑战
% In this paper, we propose to \textit{selectively transfer} KV caches for self-hosted disaggregated LLM inference systems, leveraging the sparsity in KV caches~\cite{osdi2024infinigen, nips2023h2o, iclr2025omnikv, arxiv25retroinfer, deepseek-dsa} to reduce transfer overhead and alleviate the network bottleneck.
% Specifically, only part of the important KV entries are selected and pushed to the decoding node during the prefill stage so that data transfer can be fully overlapped by the prefill computation.
% The decoding node fetches the remaining required ones from prefill nodes in an on-demand manner.
% Such a selective transfer scheme enables the decoding stage to start without waiting for the entire KV cache, ensuring a seamless stage transition.
% However, two challenges should be addressed before such an idea becomes a practical KV cache transfer solution.

% 挑战
\textbf{\textit{(1) Accurately identifying important KV entries during the prefill stage.}}
Since only a limited number of KV entries can be transferred during the prefill stage, the decoding stage has to fetch the rest of the required ones to ensure model accuracy.
The more important KV entries are identified and transferred, the fewer need to be fetched by the decoding node.
However, the state-of-the-art sparse attention algorithms~\cite{osdi2024infinigen, acl2025hata, icml2024quest, arxiv25retroinfer} depend on queries or hidden states of tokens generated during the decoding stage to define the importance of KV entries.
This makes accurate identification of important KV entries challenging in the prefill stage since the dependent information is not yet available.

\textbf{\textit{(2) Efficiently fetching KV entries during the decoding stage.}}
To ensure model accuracy, the decoding node must fetch missing KV entries from the prefill nodes before attention computations.
This process degrades decoding efficiency by adding a network round-trip to the critical path of each attention layer.
Since the remote KV fetching is required in every iteration, the cumulative overhead becomes substantial when generating a long text.

% XXX: 改成 importance aware 吧
% 为了解决上述challenge，我们提出xxx
To address the above challenges, we design \textbf{\SmartGen}, an importance-aware KV cache transfer engine that ensures
% \textbf{\underline{S}}eamless exec\textbf{\underline{\textsc{u}}}tio\textbf{\underline{\textsc{n}}} of \textbf{\underline{\textsc{d}}}is\textbf{\underline{\textsc{a}}}ggr\textbf{\underline{\textsc{e}}}gated inference for self-hosted LLM inference systems.
seamless execution of disaggregated inference for self-hosted LLM inference systems.
Specifically, \SmartGen categorizes KV cache tokens into three classes, \ie \textit{universally important}, \textit{context-dependent}, and \textit{less important}.
1) Universally important tokens are tokens in positions that are considered important regardless of input prompts. We design a \textit{profile-based proactive transfer} scheme to identify and push these tokens ahead of the decoding stage.
2) Context-dependent tokens are other tokens that are specifically important to some queries. We propose a \textit{parallel on-demand transfer} scheme to efficiently fetch these tokens during decoding, in which network round-trips overlap with local KV cache loading.
3) For the remaining less important tokens, a \textit{speculative transfer} scheme opportunistically delivers them during network idle periods, thereby amortizing cumulative transfer overhead without impacting the critical path.

We implement \SmartGen, integrate it with various sparse attention algorithms~\cite{osdi2024infinigen, acl2025hata}, and evaluate it on Llama-3.1~\cite{llama-3}, Qwen3~\cite{qwen3}, Gemma-3~\cite{gemma-3}, and Phi-4~\cite{phi-4} models with various real-world workloads~\cite{acl2024longbench, govreport, samsum, lcc}.
% with diverse real-world workloads, including question-answering~\cite{acl2024longbench}, summarization~\cite{govreport}, few-shot learning~\cite{samsum}, and code completion~\cite{lcc}.
Compared with full KV cache transfer, \SmartGen achieves up to $4.3\times$ lower TTST with similar subsequent decoding performance and accuracy.

% 总结，接贡献123
In summary, this paper makes the following contributions:
% 0. 分析并指出问题
% 1. 提出一种XXX, 解决了XXX问题
% 2. 实现并取得XXX效果
\begin{itemize}[noitemsep, topsep=0pt, parsep=0pt, partopsep=0pt]
    \item We identify the stage-transition stall issue when hosting disaggregated LLM inference systems on low-cost cloud instances, based on experimental analyses.
    \item We propose the idea of importance-aware KV cache transfer to eliminate the stall. We also address the challenges of adopting selective transfer with \SmartGen.
    \item We show that \SmartGen is a practical KV cache transfer with our experiments, achieving up to $4.3\times$ lower TTST compared with existing approaches.
\end{itemize}
\section{Background}

\subsection{Large Language Model Inference}
% Transformer-based Models
\noindent
LLMs are composed of multiple stacked transformer blocks.
Each transformer block consists of an \textit{attention layer} and a \textit{feed-forward network} (FFN)~\cite{nips2017attention}.

\textbf{\textit{Transformer computation.}}
% The input tensor $X_{in}$ to the transformer block encodes $N$ query tokens as a tensor of shape $N \times D$, where $D$ is the model dimension.
% It is then layer-normalized and fed into the attention layer.
% Within the attention layer, it is multiplied by three $D \times D$ weight matrices, \ie $W_q$, $W_k$, and $W_v$, to produce the query (Q), key (K), and value (V) matrices, each of shape $N \times D$.
% Each Q, K, and V consists of $H$ attention heads and thus are reshaped to $H \times N \times d$, where $H \times d = D$.
% Each head independently computes attention, which can be formulated as $softmax(\frac{QK^T}{\sqrt{d}})V$.
% The outputs from all heads are concatenated and projected, then passed through a residual add and layer normalization before entering the FFN.
% The FFN consists of two linear layers with an activation operation in between.
% Its output goes through another residual add to produce the transformer block's final output $X_{out}$, which has the same shape as $X_{in}$, \ie $N \times D$, enabling seamless connection to subsequent blocks.
The input tensor $X_{in} \in \mathbb{R}^{N \times D}$, representing $N$ query tokens with model dimension $D$, is first layer-normalized and fed into the attention layer.
It is multiplied by three projection matrices $W_q$, $W_k$, and $W_v \in \mathbb{R}^{D \times D}$ to produce the query, key, and value matrices, \ie $Q$, $K$, and $V \in \mathbb{R}^{N \times D}$.
Each $Q$, $K$, and $V$ consists of $H$ attention heads and are reshaped to $H \times N \times d$, where $H \times d = D$.
Each head computes attention as $softmax(\frac{QK^T}{\sqrt{d}})V$.
The outputs from all heads are concatenated and projected, then passed through a residual add and a layer normalization before entering the FFN.
The FFN consists of two linear layers with an activation operation in between.
Its output goes through another residual add to produce the final output $X_{out} \in \mathbb{R}^{N \times D}$, maintaining the shape of $X_{in}$ for the next transformer block.

% Generative Inference and KV Caches
\textbf{\textit{Generative inference and KV caches.}}
Generative LLM inference consists of two stages: \textit{prefill} and \textit{decoding}.
In the prefill stage, the LLM processes the input sequence, \ie the \textit{prompt}, to generate the first output token. %, which serves as the initial input for the decoding stage.
In the decoding stage, the LLM uses the latest token to produce the next one, forming an \textit{autoregressive} process for token generation that repeats until completion.
% To ensure that each newly generated token remains contextually coherent, 
During this process, the LLM computes the attention score of each newly generated token with all previous tokens in every iteration.
To avoid redundant computation, the $K$ and $V$ of previous tokens are cached in memory, which is known as the \textit{KV cache}.

\subsection{P/D Disaggregation}
\noindent
The use of KV caches makes the decoding stage memory-intensive, exhibiting characteristics distinct from the prefill stage.
This asymmetry leads to interference between the two stages when batched together on the same hardware.
To mitigate such interference, several works~\cite{osdi2024distserve, isca2024splitwise, icml2024dejavu, arxiv2024tetriinfer} propose to disaggregate the two stages onto separate GPUs.
To further improve resource efficiency and maximize throughput in real deployments, Mooncake~\cite{fast2025mooncake} introduces a \textit{KV-cache-centric} architecture that fully decouples prefill and decoding stages into two separated clusters and offloads KV caches in the CPU memory pool.
% 介绍RDMA原语
The prefill and decoding clusters are connected with RDMA NICs, which enable them to communicate with each other using \textit{one-sided verbs} (\eg \textsf{READ}, \textsf{WRITE}) or \textit{two-sided verbs} (\eg \textsf{SEND}, \textsf{RECV}).
However, regardless of the type of verb used, inter-node data transfer is fundamentally constrained by network bandwidth.
Given the large KV cache size, the limited RDMA bandwidth can easily become a significant performance bottleneck~\cite{isca2024splitwise, osdi2024distserve, sigcomm25hack}.
Things get worse when individuals want to deploy disaggregated inference with cloud instances, \eg EC2~\cite{aws-instances} and ECS~\cite{alibaba-cloud-instances}, since they usually offer limited network bandwidth.
HACK~\cite{sigcomm25hack} addresses network bandwidth via a 2-bit homomorphic KV cache quantization scheme that avoids dequantization overhead.
% While effective in reducing communication overhead, it can lead to noticeable accuracy trade-offs, as shown in Table~\ref{tab:hack-acc}.
In this paper, we instead reduce the network burden by exploiting the dynamic sparsity of the KV cache, which is orthogonal to quantization-based approaches.

\begin{figure}[t]
    \centering
    \includegraphics[width=\columnwidth]{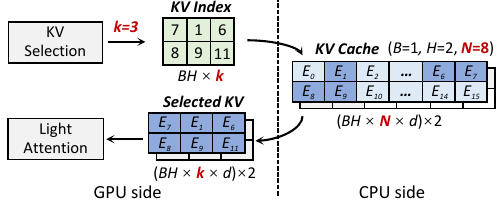}
    \caption{An example of KV selection and KV fetching from a batch of size $B$, where each head of each request selects its own top-$k$ entries. $E_i \in \mathbb{R}^{d}$ represents the $i^{th}$ KV entry.}
    \label{pic:background-kv-selection}
\end{figure}

\subsection{Dynamic KV Cache Selection}\label{sec:kv-selection}
\noindent
Due to the large size of KV caches, many studies~\cite{osdi2024infinigen, nips2023h2o, icml2024quest, arxiv2024clusterkv, iclr2024streamingllm, iclr2025omnikv, arxiv25retroinfer} propose dynamically selecting and loading only essential KV entries for attention computation, leveraging the sparsity in KV caches.
This paradigm has also been explored and adopted in industry, \eg in DeepSeek's DSA~\cite{deepseek-dsa}, Google's Spark~\cite{google-sparktrans}, and Microsoft's MInference~\cite{microsoft-minference}.
It reduces computational load, preserves model accuracy, and works well with KV cache offloading~\cite{osdi2024infinigen, icml2023flexgen, iclr2025omnikv}, making it well-suited for deployment on cloud instances with limited GPU memory.
As shown in Figure~\ref{pic:background-kv-selection}, dynamic KV selection algorithms output a \textit{KV index} matrix of shape $BH \times k$, where each row stores $k$ indexes indicating the top-$k$ KV entries selected for an attention head in a batch. 
Only the selected entries are loaded onto the GPU via host-to-device copy for attention computation.
% The selected entries are then loaded onto the GPU memory via host-to-device copy.
In this paper, we define \textit{KV selection} as identifying the KV entries to use, and \textit{KV fetching} as loading them onto the GPU.

\section{Analysis of KV Cache Transfer}\label{sec:analysis}

\noindent
This section motivates the idea (\S~\ref{sec:motivation}) and presents the challenges (\S~\ref{sec:challenges}) of adopting selective KV cache transfer for self-hosted disaggregated LLM inference.
% Without explicit mention, we use the open-source analytical model of DistServe~\cite{osdi2024distserve} to conduct simulations on NVIDIA A100-80GB GPUs.
% The simulation error is less than 2\% according to the paper.
% As for real experimental analyses, we use 4 physical machines (\ie 3 prefill nodes and 1 decoding node), with each equipped with two NVIDIA V100-32GB GPUs and a 25 Gbps Mellanox ConnectX-5 NIC.
% This network configuration aligns with those of cloud instances, \eg 25 Gbps for AWS's L4 instances~\cite{aws-instances} and Tencent Cloud's A100 instances~\cite{tencent-cloud-instances}.
All the experiments in this section are conducted with 3 \textit{ecs.gn8is-2x.8xlarge} prefill instances and one \textit{ecs.gn8is.4xlarge} decoding instance on Alibaba Cloud~\cite{alibaba-cloud-instances}. 
Each prefill instance is equipped with two NVIDIA L20 GPUs.
The decoding instance has one L20 GPU and a 25 Gbps RDMA network interface.
We offload KV caches to host memory before transfer, as GPU memory on low-cost cloud instances is limited.
The offloaded KV caches can also serve as prefix caches for reuse, like existing P/D disaggregated systems~\cite{fast2025mooncake, nvidia-dynamo}.
We assume a prefix cache hit ratio of 75\% in host memory, reflecting a representative value across reported hit ratios ranging from 60\% to 90\% in various long-context scenarios~\cite{arxiv2025cachewild, fast2025mooncake, nips2024sglang, atc2024cachedattention, eurosys25pensieve}.

\subsection{Motivation: Issues with Full Transfer}\label{sec:motivation}
\noindent
% Modern disaggregated LLM inference systems transfer KV caches via high-speed RDMA networks.
% However, the enormous KV cache could result in a significant transfer overhead~\cite{osdi2024distserve, isca2024splitwise}.
% Prior studies have shown that KV cache transfer over PCIe already becomes a bottleneck~\cite{osdi2024infinigen}, let alone the cross-node RDMA network with even lower bandwidth on rented cloud instances. 
% 强调motivation
In disaggregated inference systems, the KV cache transfer overhead directly affects the delivery of the second token to users, making TTST as important as time-to-first-token (TTFT).
% Specifically, LLM serving systems typically use an output buffer to store generated but not yet consumed tokens~\cite{eurosys26tokenflow}. 
% If the buffer empties before new tokens are available, users experience stalls.
% Poor TTST hinders the buffer from generating enough tokens in time to return to the user.
Specifically, LLM serving systems use an output buffer to store generated tokens and deliver them progressively to users~\cite{eurosys26tokenflow}.
If the buffer empties before new tokens are available, users experience stalls.
Since the first token is sent to the user almost immediately, a long delay in producing the second token leaves the buffer empty, resulting in a user stall.
Existing systems~\cite{isca2024splitwise, arxiv2024tetriinfer, icml2024dejavu, fast2025mooncake, nvidia-dynamo} hide KV cache transfer latency by overlapping it with prefill computation.
However, our observations show that the overlapping strategy fails to fundamentally address the bandwidth bottleneck. %, particularly in large-scale deployments.

\begin{figure}[t]
    \vspace{-3mm}
    \footnotesize
    \centering
    \subfloat[Varying model sizes.]{
        \includegraphics[width=0.524\columnwidth]{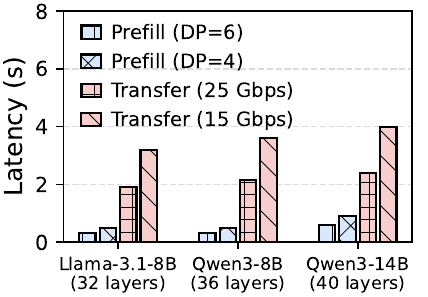}
        \label{fig:transfer-model-size}
    }
    \hspace{-0.345cm}
    \subfloat[Varying batch sizes.]{
        \includegraphics[width=0.476\columnwidth]{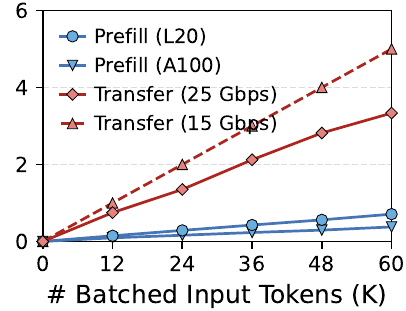}
        \label{fig:transfer-batch-size}
    }
    \caption{Comparison of the full KV cache transfer overhead and the prefill computation latency on Llama and Qwen models.}
    \label{fig:motivations}
\end{figure}

\noindent
\textbf{\textit{\underline{Observation}: KV cache transfer cannot be overlapped by prefill computation.}}
In the P/D disaggregation architecture, the compute-intensive prefill nodes typically adopt small batch sizes to satisfy TTFT SLOs~\cite{osdi2024distserve, isca2024splitwise, arxiv2024tetriinfer}.
Multiple prefill nodes are often adopted to handle highly concurrent user requests.
On the decoding side, due to the memory-intensive nature of the decoding stage, batch sizes of decoding nodes are always larger than those of prefill nodes~\cite{osdi2024distserve, isca2024splitwise, arxiv2024tetriinfer, fast2025mooncake}.
Consequently, KV caches have to be transferred from multiple prefill nodes to a single decoding node, saturating the decoding node's limited network bandwidth.
% The problem becomes even worse as the model size and sequence length grow.
The problem worsens as model size and sequence length grow.

Figure~\ref{fig:transfer-model-size} shows the prefill and KV cache transfer latency of 48K batched input tokens under three different models, respectively.
Larger models generate larger KV caches, and as a result, the KV cache transfer latency increases by $1.3\times$.
In contrast, the prefill latency remains below 0.6 seconds with a data parallelism (DP) degree of 6.
This makes it hard for the prefill stage to hide the KV cache transfer as the model size increases.
Figure~\ref{fig:transfer-batch-size} shows the results on the Qwen3-14B model with a DP degree of 6 and various numbers of batched input tokens.
As the number of tokens increases, the gap between transfer latency and prefill latency on our testbed reaches up to $4.7\times$, indicating a more severe stage-transition stall. 
If the network bandwidth is further limited to 15 Gbps (\eg using a smaller L20 instance, \textit{ecs.gn8is.2xlarge}~\cite{alibaba-cloud-instances}, for decoding), this gap can be amplified up to $7.1\times$.

% \textbf{\textit{\underline{Observation 2}: Transfer remains under-optimized relative to prefill.}}
% \noindent
% \textbf{\textit{\underline{Observation 2}: Prefill optimizations make it harder to overlap KV cache transfer.}}
% Due to the time-to-first-token (TTFT) SLO, the prefill stage has received extensive attention and optimizations, \eg FlashAttention~\cite{nips2022flashattention} and prefix caching~\cite{nips2024sglang, fast2025mooncake, sosp2023vllm, fast2025impress}.
% Additionally, prefill optimizations make it harder to overlap the KV cache transfer.
% The prefill stage has received extensive optimizations, \eg FlashAttention~\cite{nips2022flashattention} and prefix caching~\cite{nips2024sglang, fast2025mooncake, sosp2023vllm, fast2025impress}.
% It is reported that the prefix caching in Mooncake can reduce prefill latency by 86\% with a prompt length of 128K tokens~\cite{fast2025mooncake}.
% Besides, improved hardware capabilities have also boosted the prefill efficiency.

If high-end GPUs are adopted during the prefill stage, it would be even harder to overlap the KV cache transfer with the prefill computation due to the higher execution efficiency.
% prefill computation is further accelerated, making it even harder to overlap the KV cache transfer.
% Figure~\ref{fig:transfer-batch-size} shows the simulated result of A100 GPUs using DistServe~\cite{osdi2024distserve}'s analytical model.
% The simulation error is less than 2\% according to the paper.
As shown in Figure~\ref{fig:transfer-batch-size}, with more Tensor Cores and higher VRAM bandwidth, A100 achieves up to a $1.9\times$ speedup over L20 when prefilling 60K tokens, making the KV transfer issue more pronounced.
% Moreover, advanced prefill optimizations, \eg prefix caching, would make the problem worse.
% It is reported that prefix caching can further reduce prefill latency by 86\% when the prompt length is 128K tokens~\cite{fast2025mooncake}.

% In contrast, the optimization of KV cache transfer remains insufficient and even suffers from network congestion.
% When massive KV caches saturate the RDMA bandwidth, excess data accumulates in the network, causing queuing delays.
% As shown in Figure~\ref{fig:network-congestion}, the divergence (\ie shown as shadow) between the actual and theoretical transfer latency indicates the impact of network congestion.
% % With a batch size of 30, the transfer latency is amplified by $1.9\times$. 
% % This further worsens the KV cache transfer issue.
% The transfer latency is amplified by up to $1.6\times$, worsening the transfer issue.

% （弃）下面这段或许可以改成idea是跨两阶段的KV cache transfer
% root cause: decoding存在启动约束：必须要等待所有KV cache到达后才能开始，所以只能用prefill来掩盖
% idea: 用KV sparsity来break上述约束，以此让KV传输被两阶段一起掩盖
\noindent
\textbf{\textit{\underline{Opportunity}: Not all KV entries are equally important.}}
Our work is inspired by the KV cache sparsity, \ie selecting only the most critical tokens' KV caches for attention can maintain comparable model accuracy~\cite{osdi2024infinigen, nips2023h2o, icml2024quest, arxiv2024clusterkv, iclr2025omnikv, iclr2024streamingllm, acl2025hata}.
Based on this, we propose to address the network bandwidth bottleneck with \textbf{\textit{selective KV cache transfer}}.
% Specifically, we adopt InfiniGen~\cite{osdi2024infinigen}, the state-of-the-art KV cache selection algorithm, to exploit KV cache sparsity.
Specifically, one of the existing dynamic KV cache selection algorithms~\cite{osdi2024infinigen, acl2025hata} is adopted to exploit KV cache sparsity.
During the prefill stage, only part of the essential KV entries are selected and pre-transferred to the decoding node, with the hope that the KV cache transfer can be fully overlapped by the prefill computation.
During the decoding stage, to maintain model accuracy, the decoding node fetches any missing KV entries from prefill nodes in an on-demand manner.
% \footnote{In this paper, we choose to adopt InfiniGen~\cite{osdi2024infinigen}, the state-of-the-art dynamic KV selection algorithm, on the decoding node.}
% This selective scheme amortizes the KV cache transfer across the prefill and decoding stages, making it possible to fully hide the transfer overhead.

\subsection{Challenges of Selective Transfer}\label{sec:challenges}
% 引入challenge
% Although selective KV cache transfer could relieve the network burden, it introduces new challenges due to \textit{KV selection} during the prefill stage and \textit{KV fetching} during the decoding stage.
\noindent
Although selective KV cache transfer could relieve the network burden, it introduces new challenges due to the \textit{KV selection} and \textit{KV fetching} during the two stages, respectively.

\begin{figure}[t]
    \vspace{-1mm}
    \footnotesize
    \centering
    \subfloat[KV selection.]{
        \includegraphics[width=0.505\columnwidth]{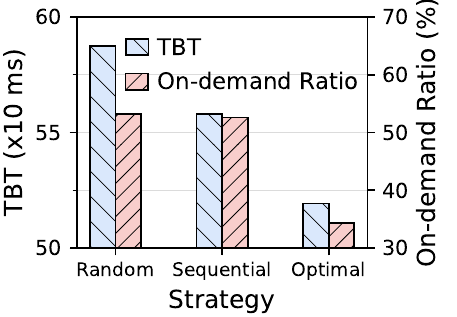}
        \label{fig:selection-opt-space}
    }
    \hspace{-0.365cm}
    \subfloat[KV fetching.]{
        \includegraphics[width=0.495\columnwidth]{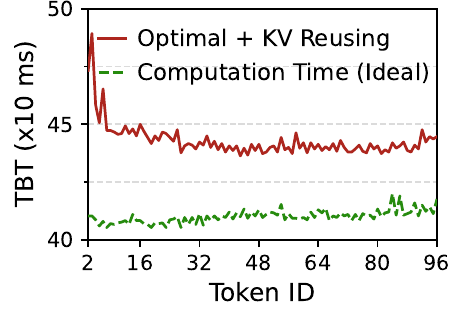}
        \label{fig:offloading-opt-space}
    }
    \caption{Optimization opportunities in prefill-side KV selection and decoding-side KV fetching on the Qwen3-14B model, assuming that half of KV cache can be transferred during the prefill stage.}
    \label{fig:challenges}
\end{figure}

\noindent
\textbf{\textit{\underline{Challenge 1}: Accurate KV selection on the prefill node.}}
The accuracy of KV selection during the prefill stage directly impacts the overhead of on-demand KV fetching during the decoding stage.
% Transferring more relevant KV entries upfront reduces the number of entries that must be fetched later by the decoding node.
% Figure~\ref{fig:selection-opt-space} shows the impact of such KV selection on the time-between-tokens (TBT) in the decoding stage with 12K batched input tokens.
We define the \textit{on-demand ratio} as the ratio of KV entries fetched on demand from the prefill node to the total KV entries required by the decoding node.
Transferring more relevant KV entries upfront reduces the ratio, thereby reducing the time-between-tokens (TBT) in the decoding stage.
Figure~\ref{fig:selection-opt-space} shows the impact of such KV selection with 48K batched input tokens.
% The vertical line represents the theoretical maximum percentage of the KV cache that can be transferred within the prefill computation time according to Figure~\ref{fig:transfer-model-size}.
With an optimal selection strategy, the most critical KV entries are selected and pre-transferred during the prefill stage.
This strategy reduces the on-demand ratio from 53\% to 34\% and optimizes TBT by $1.1\times$, compared with the typical sequential strategy where KV entries are transferred in memory address order.
A similar observation holds when compared with a random strategy.

However, achieving the optimal selection strategy is challenging since state-of-the-art KV selection methods are dynamic.
They rely on query~\cite{icml2024quest, arxiv2024clusterkv, iclr2025omnikv, acl2025hata} or hidden states~\cite{osdi2024infinigen} generated during the decoding stage to accurately identify important KV entries.
Since the decoding stage has not yet started during the prefill stage, the prefill node has no visibility into which KV entries the decoding node will select.

\noindent
\textbf{\textit{\underline{Challenge 2}: Efficient KV fetching on the decoding node.}}
To ensure decoding accuracy, the dynamically selected KV entries must be locally available before the transformer computation.
This requires the decoding node to check and fetch any missing entries from prefill nodes before loading them onto the GPU, introducing a network round-trip on the critical path of the host-to-device copy.
% As shown in Figure~\ref{fig:selection-opt-space}, even if the maximum percentage of KV cache that the prefill computation can overlap is transferred, remote KV fetching still incurs at least a $1.3\times$ increase in TBT compared with the case when decoding computation does not need to fetch remote KV caches, \ie the ideal case.
% As shown in Figure~\ref{fig:selection-opt-space}, even if the maximum percentage of KV cache that the prefill computation can overlap is transferred, there still remain 60\% of KV entries that must be fetched remotely.
As shown in Figure~\ref{fig:selection-opt-space}, even if half of the KV cache is pre-transferred, there still remain 34\% of KV entries that must be fetched remotely.

Worse still, inefficient KV fetching persists throughout the decoding stage.
Although later tokens can reuse previously fetched KV entries, the fine-grained dynamic selection always results in new missing entries that must be fetched from prefill nodes.
% As shown in Figure~\ref{fig:offloading-opt-space}, despite reusing previously fetched entries, the overhead of remote KV fetching persists in subsequent decoding iterations, resulting in a sustained $1.1\times$ increase in TBT.
% This overhead accumulates with the output length, leading to a high end-to-end latency.
As shown in Figure~\ref{fig:offloading-opt-space}, remote KV fetching still incurs a sustained $1.1\times$ TBT increase in each decoding step even if previously fetched KV entries are reused.
% This overhead accumulates over long outputs, resulting in high end-to-end latency.

\section{The \SmartGen Design}\label{sec:design}

\begin{figure}[t]
    \centering
    \includegraphics[width=\columnwidth]{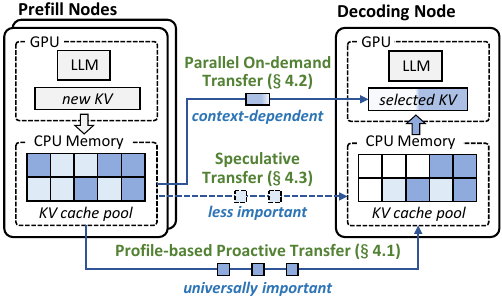}
    \caption{The overview of \SmartGen.}
    \label{pic:overview}
\end{figure}

\noindent
We propose \SmartGen, an importance-aware KV cache transfer engine that enables seamless state transitions in disaggregated inference for self-hosted LLMs on the cloud.
As shown in Figure~\ref{pic:overview}, \SmartGen categorizes KV cache entries into three types and comprises three transfer paths.
First, to achieve accurate KV selection during the prefill stage, \SmartGen adopts a \textit{profile-based proactive transfer} to push the universally important KV entries to the decoding node (\S~\ref{sec:proactive-transfer}).
Second, to achieve fast KV fetching during the decoding stage,  \SmartGen conducts a \textit{parallel on-demand transfer} to fetch missing context-dependent KV entries, which overlaps the network overhead with local KV cache loading (\S~\ref{sec:on-demand-transfer}).
Finally, \SmartGen proposes a \textit{speculative transfer} to deliver all remaining less important KV entries to the decoding node (\S~\ref{sec:speculative-transfer}).

%%%%%%%%%%%%%%%%%%%%%%%%%%%%%%%%%%%%%%%%%%%%%%%%%%%%%%%%%%%
\subsection{Profile-based Proactive Transfer}\label{sec:proactive-transfer}
% profile-based proactive是用来解决XXX => 关键挑战是XXX => 我们提出使用positional similarity（需要解释）来解决
\noindent
Profile-based proactive transfer is presented to eliminate the stage-transition stall by transferring only essential KV entries.
The key challenge lies in accurately identifying essential KV entries during the prefill stage.
% Many static KV sparsity algorithms~\cite{iclr2024streamingllm, iclr25duoattn, deepseek-nsa} have demonstrated the model and workload stability of static KV importance.
Inspired by static KV sparsity~\cite{iclr2024streamingllm, iclr25duoattn, deepseek-nsa}, we first propose to address this challenge in this section by exploiting \textit{positional similarity}, \ie universally important KV entries tend to appear at similar positions in input sequences.
We then describe how we use the positional similarity to guide the accurate selection and efficient RDMA-based transfer of essential KV entries. 

\begin{figure}[t]
    \vspace{3mm}
    \centering
    \includegraphics[width=1.0\linewidth]{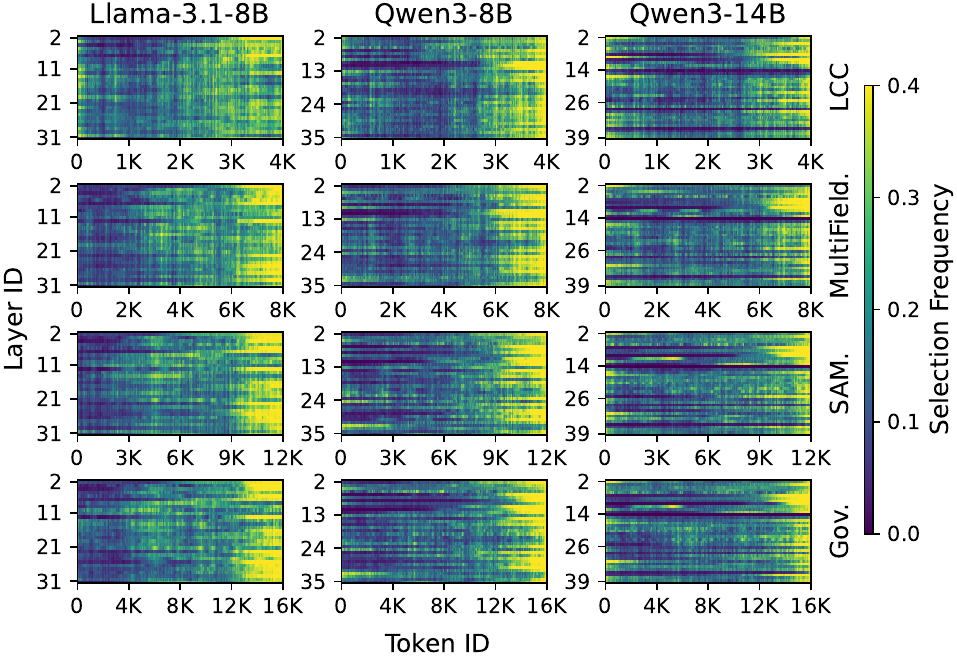}
    \caption{The frequency of selecting each token in each attention layer on various models~\cite{llama-3, qwen3} and datasets~\cite{acl2024longbench, lcc, samsum, govreport}.}
    \label{fig:profiling}
\end{figure}

% insight (重要kv的位置相似性) + offline阶段 (profile质量+数量) + online阶段 (rdma、KV mask)
\textbf{\textit{Positional similarity in important KV entries.}}
% \subsubsection{Positional similarity in important KV entries.}
% Dynamic KV selection algorithms~\cite{osdi2024infinigen, icml2024quest, nips2023h2o} generally use the magnitude of attention scores as a metric to evaluate KV importance.
% Without loss of generality, we use InfiniGen~\cite{osdi2024infinigen} to demonstrate the positional frequency distribution of important KV entries in a batch of size 6, as shown in Figure~\ref{fig:profiling}.
We define a KV entry's importance as how frequently it is selected by the KV selection algorithm during the decoding stage.
Figure~\ref{fig:profiling} shows the importance distribution of KV entries selected by InfiniGen~\cite{osdi2024infinigen}. 
The results can be applied to other selection algorithms~\cite{acl2025hata, icml2024quest, nips2023h2o, arxiv25retroinfer} since they all use the magnitude of attention scores to select KV.
% This definition also applies to other KV selection algorithms, as existing algorithms~\cite{osdi2024infinigen, icml2024quest, nips2023h2o} generally use the magnitude of attention scores to evaluate KV importance.
% Without loss of generality, Figure~\ref{fig:profiling} presents the importance distribution of important KV entries selected by InfiniGen~\cite{osdi2024infinigen} in a batch of size 6.
To simplify comparison, we truncate requests in various datasets~\cite{acl2024longbench, lcc, samsum, govreport} to 4K, 8K, 12K, and 16K, respectively.
% 分析实验结果
% 1. 同一model在不同数据集下的KV重要性分布很相似 => 说明了profile的可行性
% 2. 这样的数据集无感的分布在不同的模型下也存在 => 说明了profile的普适性
% Our profile-based method builds on the key observation that \textit{important KV entries exhibit similar relative positional patterns across different datasets}.
Our key observation is that \textit{important KV entries tend to appear in consistent regions of the KV cache across different datasets}.
% As illustrated in Figures~\ref{fig:profile-opt-13b-longbench} and~\ref{fig:profile-opt-13b-sharegpt}
For the same model, datasets with different prompt lengths do not affect the relative positional distribution of important KV entries.
For instance, in attention layer 8 of Qwen3-14B, each of the last 25\% of tokens is selected by over 30\% of heads and requests, regardless of the prompt length, while in layer 14, most tokens are not selected by the majority of heads and requests.
Similar patterns are also observed in other models of different sizes and architectures, \eg Llama-3.1-8B.
This indicates that, within a certain range of prompt lengths, \eg 4K-16K, a calibration dataset can approximate the token-importance distribution for requests in that range.

% As shown in Figure~\ref{fig:positional-similarity}, we quantify positional similarity by computing the Jensen–Shannon (JS) divergence between selection distributions across different datasets, and then averaging the results over attention layers.
% The JS divergence increases slowly as the difference in prompt lengths grows, indicating that, within a certain range of prompt lengths, \eg 4K-16K, a calibration dataset can approximate the token-importance distribution for requests in that range.

\textbf{\textit{Offline KV selection.}}
% 切块 + profile (质量公式、数量公式) + 全局top-k
Based on the above observation, we conduct \textit{offline profiling} on ranges of prompt lengths to 1) identify important regions in KV cache matrices and 2) help predict how many KV entries can be transferred during the prefill computation.
Without loss of generality, the following describes a single profiling run.

First, we run a round of model inference on a calibration dataset to calculate the frequency of selecting each KV entry.
Then, for each layer, we partition the KV cache along the sequence dimension into $M$ KV blocks.
In this way, we can use fixed-sized KV blocks to approximate variable-length regions.
Thus, there are in total $L \cdot M$ KV blocks, where $L$ is the number of attention layers.
Let $B_{l,m}$ denote the $m$-th KV block in layer $l$.
% The importance $I_{l,m}$ of the KV region corresponding to block $B_{l,m}$ is defined as the average frequency: 
The importance $I_{l,m}$ of block $B_{l,m}$ is defined as the average frequency: 
\begin{equation}\label{eq:importance}
I_{l,m} =
\begin{cases}
\infty, & \text{if } l = 0 \text{ or } 1 \\
\frac{1}{|E(B_{l,m})|}\sum_{e \in E(B_{l,m})}S(e), & \text{if } 2 \le l < L
\end{cases}
\end{equation}
% \begin{equation}\label{eq:importance}
% I_{l,m} =
% \begin{cases}
% \infty, & \text{if } l = 0 \text{ or } 1 \\
% \frac{1}{|T(B_{l,m})|}\sum_{t \in T(B_{l,m})}S(t), & \text{if } 2 \le l < L
% \end{cases}
% \end{equation}
Here, $E(B_{l,m})$ is the set of KV entries in KV block $B_{l,m}$, and $S(e)$ denotes the number of times entry $e$ is selected in the decoding stage during the offline profiling.
KV blocks with higher importance are prioritized for selection and transfer during the online prefill stage.
The KV caches of the first two attention layers, \ie $l=0,1$, are always selected as they are consistently important and required by the decoding stage~\cite{osdi2024infinigen, icml2024quest, arxiv2024clusterkv}.
We set $M$ to 1K in our implementation.
% , balancing the accuracy of KV selection and the capacity of RDMA NICs.

\begin{figure}[t]
    \vspace{3mm}
    \centering
    \includegraphics[width=\columnwidth]{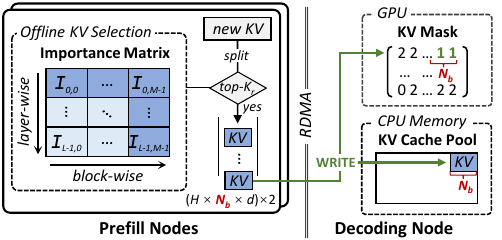}
    \caption{The process of profile-based proactive transfer, where $N_b$ is the number of tokens in the KV block. The KV mask is used for on-demand transfer, which will be introduced in \S~\ref{sec:on-demand-transfer}.}
    \label{pic:proactive-transfer}
\end{figure}

In addition, we also profile the prefill computation time and full KV cache transfer time to know how many KV blocks can be transferred during prefill computation.
Based on the profiled prefill time $T_p$ and transfer time $T_t$, we select the top-$K_r$ most important KV blocks among all $L$ layers, where:
\begin{equation}\label{eq:top-k}
K_{r} = L \cdot M \cdot \left( \frac{L - 1}{L} \cdot T_p / T_t \right) = M \cdot (L - 1) \cdot \frac{T_p}{T_t}
\end{equation}
Here, $K_{r}$ is calculated as the product of the total number of KV blocks and the proportion of transfer time that can be overlapped with prefill computations.
As KV cache transfer can only start after the first attention layer finishes computation, we consider only the prefill duration of $L-1$ layers, \ie $\frac{L-1}{L}\cdot T_p$.
The estimation provides an upper bound on the number of KV blocks whose transfer can be fully overlapped with the prefill computation of all but the first transformer block.
In practice, we clip $K_r$ between $2M$ and $LM$ to ensure that KV blocks of the first two layers are always selected.

\textbf{\textit{Online KV transfer.}}
Figure~\ref{pic:proactive-transfer} shows the process of profile-based proactive transfer.
Each time the prefill stage offloads an attention layer's KV cache to CPU memory, \SmartGen first splits it into $M$ blocks.
For each KV block, it checks whether its position belongs to the top-$K_r$ most important ones.
If so, \SmartGen uses one-sided RDMA \textsf{WRITE} to push the block to the decoding node's KV cache pool.

% To enable the decoding node to track the status of each KV entry, we pre-allocate an \textit{int8} \textit{KV mask} matrix of shape $B \times N$ for each attention layer on the decoding node, where $B$ is the batch size and $N$ is the sequence length. 
% \footnote{The memory overhead of KV mask matrices is negligible.
% When serving 64K tokens on Qwen3-14B, the overhead is only $L \cdot 64$ KB $= 2.5$ MB.}
% Each element $m_{ij}$ denotes the status of the $j^{th}$ token in the $i^{th}$ sequence.
% All heads of the same token share a mask value for their KV entries.
% A value of 0 indicates that the corresponding KV entries belong to a padded token in the batch.
% The padded tokens are used to align input sequences across the batch, and the KV mask value of 0 prevents their KV caches from being loaded onto the GPU.
% A value of 1 indicates that the KV entries are locally available, and 2 indicates that they reside on a remote prefill node.
% Therefore, each time a KV block is pushed to the decoding node, the prefill node also updates the corresponding values in the decoding-side KV mask via an additional RDMA \textsf{WRITE}.
% The two \textsf{WRITE}s are combined into a single network round-trip using doorbell batching.
% This leverages the in-order delivery property of RDMA NICs~\cite{infiniband-arch-spec, vldb2019in-order} to ensure that the KV mask is updated only after the corresponding KV block has been written.

\begin{figure}[t]
    \centering
    \includegraphics[width=\columnwidth]{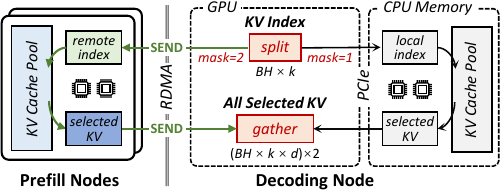}
    \caption{The process of parallel on-demand transfer.}
    \label{pic:on-demand-transfer-w-gdr}
\end{figure}

%%%%%%%%%%%%%%%%%%%%%%%%%%%%%%%%%%%%%%%%%%%%%%%%%%%%%%%%%%%
\subsection{Parallel On-demand Transfer}\label{sec:on-demand-transfer}
\noindent
% Although profile-based proactive transfer can pre-transfer important KV entries, the limited number of transferable KV blocks means that the decoding node still needs to fetch missing KV entries on demand.
% This takes place on the critical path of local KV cache loading, degrading decoding efficiency.
% This section introduces a two-lane KV fetching technique to remove the network round-trip from the critical path.
% The following starts with cloud instances supporting GDR.
The limited number of transferable KV blocks still forces the decoding node to fetch missing entries on demand, placing network round-trips on the critical path.
To address this, we introduce a two-lane KV fetching technique that removes such round-trips from the critical path.

% \textbf{\textit{Two-lane parallel KV fetching.}}
% 引入KV index是出发点
Our key idea is to parallelize the loading of local and remote KV caches.
As introduced in Section~\ref{sec:kv-selection}, the KV index generated by the KV selection algorithm determines KV entries to load onto the GPU. 
We split the KV index into two separate indexes, \ie one for local and the other for remote KV entries, so as to \textit{decouple the KV cache loading into two parallel data transfer lanes}, as shown in Figure~\ref{pic:on-demand-transfer-w-gdr}.
% 介绍two lanes
With the split index, the decoding node selects and fetches KV entries from host memory using only the local index.
At the same time, it issues remote procedure calls (RPCs) to prefill nodes to fetch the missing entries identified by the remote index.
% 介绍rdma流程
In particular, it \textsf{SEND}s the remote index from its GPU memory to the host memory of prefill nodes via GPU-direct RDMA (GDR).
After \textsf{RECV}ing the index, the prefill node selects the required KV entries locally and \textsf{SEND}s them back to the decoding node's GPU memory.
The decoding node starts the attention computation once it \textsf{RECV}s all remotely fetched entries and finishes loading the local ones.

% 引入两个细节挑战
% This technique reduces KV cache loading time on the decoding node by enabling the remote and local KV fetching to proceed in parallel.
% The main challenges lie in how to efficiently split the on-GPU KV index and how to automatically gather all selected KV entries received from the two lanes:
% Specifically, for index splitting, the key issue is that the on-GPU KV index must identify which KV entries reside in host memory, despite lacking direct visibility into it.
The main challenges are efficiently splitting the on-GPU KV index and automatically gathering selected KV entries from both lanes:
1) For index splitting, since KV caches are transferred to CPU memory in a fine-grained and non-contiguous manner, the GPU cannot efficiently determine whether each KV entry is local or not.
2) For KV gathering, the dynamic and discrete positions of missing KV entries make it difficult for the RDMA NIC to place remotely fetched entries directly into the correct locations in GPU memory.

% \begin{figure}[t]
%     \centering
%     \includegraphics[width=\columnwidth]{Pictures/on-demand-transfer.pdf}
%     \caption{The process of parallel on-demand transfer.}
%     \label{pic:on-demand-transfer}
% \end{figure}

% \begin{figure}[t]
%     \centering
%     \includegraphics[width=\columnwidth]{Pictures/on-demand-transfer-wo-gdr.pdf}
%     \caption{A workaround for instances lacking GDR.}
%     \label{pic:on-demand-transfer-wo-gdr}
% \end{figure}

% \begin{figure}[t]
%     \centering
%     \subfloat[For cloud instances supporting GDR.]{
%         \includegraphics[width=\columnwidth]{Pictures/on-demand-transfer.pdf}
%         \label{pic:on-demand-transfer-w-gdr}
%     }
%     \subfloat[For cloud instances lacking GDR.]{
%         \includegraphics[width=\columnwidth]{Pictures/on-demand-transfer-wo-gdr.pdf}
%         \label{pic:on-demand-transfer-wo-gdr}
%     }
%     \caption{The process of parallel on-demand transfer.}
%     \label{pic:on-demand-transfer}
% \end{figure}

\SetKwInOut{KwIn}{Input}
\SetKwInOut{KwOut}{Output}

\definecolor{darkgreen}{RGB}{0, 102, 5}
\newcommand{\myCmSty}{\color{darkgreen}\itshape\sffamily\small}
\SetCommentSty{myCmSty}

\newcommand{\circlemark}[3]{\makebox[0.88\dimexpr\linewidth-#1\algomargin\relax][l]{#2\hfill\textcircled{\footnotesize#3}}}

\DontPrintSemicolon

\begin{algorithm}[t]
    \small
    \KwIn{
        $kv\_index$: The KV index of shape $BH \times k$ \\
        $kv\_mask$: The KV mask of shape $B \times N$
    }
    \KwOut{
        $local\_index$: The local KV index array \\
        $remote\_index$: The remote KV index array
    }

    % \medskip
    \tcc{Select required mask values from the KV mask matrix}
    \circlemark{0}{$mask \leftarrow kv\_mask \left[ kv\_index \right] $}{1}\;
    \If(\tcc*[f]{For fast KV gathering}){enable reordering}{
        \circlemark{1.8}{$mask, order \leftarrow SortEachRow(mask) $}{4}\;
        \circlemark{1.8}{$kv\_index \leftarrow kv\_index \left[ order \right] $}{5}\;
    }
    \tcc{Mask values of 1 (2) indicate local (remote) KV entries}
    \circlemark{0}{$local\_index \leftarrow kv\_index \left[ mask = 1 \right] $}{2}\;
    \circlemark{0}{$remote\_index \leftarrow kv\_index \left[ mask = 2 \right] $}{3}\;
    \Return $local\_index$, $remote\_index$\;

    \caption{\small KV Index Splitting and Reordering}
    \label{alg:kv-index-split}
\end{algorithm}

\textbf{\textit{Mask-based index splitting.}}
To address the first challenge, our key idea is to \textit{maintain mask matrices on the GPU to track the status of each KV entry}.
This enables the decoding node to observe real-time KV status in host memory directly on the GPU execution path.
Specifically, we pre-allocate an \textit{int8} KV mask matrix for each attention layer and have prefill nodes update it via GDR.
Each KV mask matrix is of shape $B \times N$, where $B$ is the batch size and $N$ is the sequence length. 
\footnote{The memory overhead of KV mask matrices is negligible.
When serving 64K tokens on Qwen3-14B, the overhead is only $L \cdot 64$ KB $= 2.5$ MB.}
Each element $m_{ij}$ denotes the status of the $j^{th}$ token in the $i^{th}$ sequence.
All heads of the same token share a mask value for their KV entries.
A value of 0 indicates that the corresponding KV entries belong to a padded token in the batch.
The padded tokens are used to align input sequences across the batch, and the KV mask value of 0 prevents their KV caches from being loaded onto the GPU.
A value of 1 indicates that the KV entries are locally available, and 2 indicates that they reside on a remote prefill node.
Therefore, each time a KV block is pushed to the decoding node, the prefill node also updates the corresponding values in the decoding-side KV mask via an additional RDMA \textsf{WRITE}, as shown in Figure~\ref{pic:proactive-transfer}.
The two \textsf{WRITE}s are combined into a single network round-trip using doorbell batching.
This leverages the in-order delivery property of RDMA NICs~\cite{infiniband-arch-spec, vldb2019in-order} to ensure that the KV mask is updated only after the corresponding KV block has been written.

Based on this, we extend the on-GPU KV selection process with several operators to split the generated KV index into local and remote ones, as shown in Algorithm~\ref{alg:kv-index-split}.
The operator \textcircled{\footnotesize 1} selects the status of the required KV entries from the KV mask matrix, generating a small mask matrix of the same shape as the KV index, \ie $BH \times k$.
The local and remote indexes are then generated by retrieving the KV index with the corresponding mask value, \ie operators \textcircled{\footnotesize 2} and \textcircled{\footnotesize 3}, where 1 indicates that the corresponding index refers to a local KV entry and 2 indicates a remote one. 
% 总结好处
These three operators efficiently split the KV index.
They involve only lightweight indexing operations, \eg indexing into the KV mask of shape $B \times N$ with a small KV index of shape $BH \times k$.
% They could be integrated into the KV selection kernel, avoiding launching an extra one.
% They could be appended directly after the KV selection code with minimal engineering overhead.
% They could be executed directly after the KV selection phase with minimal engineering overhead.
They could be easily applied to existing KV selection processes~\cite{osdi2024infinigen, acl2025hata} without modifying their code.

\textbf{\textit{Reordering-based KV gathering.}}
On the remote lane, we let prefill nodes directly send the selected KV entries to their target locations in the GPU memory of the decoding node.
This could avoid the need for the decoding node to launch an additional kernel on the critical path to scatter the received entries into the locally fetched KV cache.
However, since the target locations of remote KV entries are interleaved among local ones, discretely transferring them would incur excessive I/O overhead, which is impractical.

To address the above challenge, we leverage the fact that \textit{the order of KV entries along the sequence dimension in the KV cache does not affect the attention result}, \ie $softmax(\frac{QK^T}{\sqrt{d}})V$.
% This is because the positional information has already been encoded into the KV cache in advance.
This is because the KV cache already contains the positional information.
% Based on this, we propose reordering KV entries along the sequence dimension to make remote ones contiguous in each row of the KV cache.
Based on this, we propose reordering KV entries to make the remote ones contiguous along the sequence dimension.
% For clarity, we define a \textit{row} of KV entries as those along the sequence dimension for a given head, \ie the $k$ in shape $BH \times k \times d$.
For clarity, we define a \textit{row} of KV entries as the $k$ entries for a given head along the sequence dimension, \ie the $k$ in shape $BH \times k \times d$.
Since the order of selected KV entries is determined by the KV index, we realize the reordering with two additional operators in Algorithm~\ref{alg:kv-index-split}.
% \begin{subequations}
% \begin{align}
% mask, order & := SortEachRow(mask) \label{eq:sort-mask} \\
% kv\_index & := kv\_index \left[ order \right] \label{eq:sort-index}
% \end{align}
% \end{subequations}
The operator \textcircled{\footnotesize 4} sorts each row of the index mask in descending order, and the operator \textcircled{\footnotesize 5} applies the ordering to the KV index. 
As the highest mask value (\ie 2) indicates a remote KV entry, the descending order makes the remote KV entries contiguous at the start of each row of the selected KV cache.
% Since the highest mask value (\ie 2) indicates a remote KV entry, the descending order makes the remote KV entries contiguous at the beginning of the sequence dimension in the selected KV cache.
% Therefore, the decoding node can \textsf{RECV} these entries row by row via GDR.
Therefore, the decoding node can \textsf{RECV} these entries row-wise instead of individually via GDR.

% scatter-gather
To further accelerate the multi-row transfer, we utilize the scatter capacity of the DMA engine in the RDMA NIC~\cite{eurompi2016scatter-gather, hotos21zerocopy}, which enables the decoding node to receive a contiguous chunk of KV entries and scatter them into discrete rows at low runtime costs.
% Although the RDMA NIC's scatter-gather capability has an upper limit (\eg 20), it sufficiently reduces the number of I/Os required to transfer KV entries of $BH$ rows to a reasonable order of magnitude.
Although the scatter-gathering capacity of the RDMA NIC has an upper limit (\eg $20$), it is sufficient to significantly reduce the number of I/Os (\eg by $20 \times$).
% (\eg to $\frac{BH \times 2}{20}$)

% \textbf{\textit{Supporting cloud instances lacking GDR.}}
% During system deployment, we found that many low-cost cloud servers do not support GDR due to virtualization constraints.
% In such environments, the previous design requires two modifications, as shown in Figure~\ref{pic:on-demand-transfer-wo-gdr}:
% \textit{(1) On-CPU index splitting.}
% Since prefill nodes cannot update the KV mask matrix via GDR, the KV mask should be maintained in host memory. Consequently, the KV index is offloaded to host memory before being split.
% \textit{(2) Extra local KV loading.}
% Since the selected remote KV entries cannot be directly transferred into the GPU, the decoding node should first receive them in host memory and then load them into GPU via an extra host-to-device copy.
% Given that PCIe bandwidth is much higher than the network bandwidth, these overheads are acceptable.

%%%%%%%%%%%%%%%%%%%%%%%%%%%%%%%%%%%%%%%%%%%%%%%%%%%%%%%%%%%
\subsection{Speculative Transfer}\label{sec:speculative-transfer}
\noindent
Despite the efficiency of parallel on-demand transfer, it is still difficult to fully hide the remote lane's network overhead.
The overhead can accumulate across decoding iterations, increasing end-to-end latency.
This section eliminates the overhead for later iterations by speculatively delivering all KV entries to the decoding node in the background.
However, the background traffic may interfere with the foreground on-demand transfers since it can contend for the limited network and compute resources.
% This process runs asynchronously in the background, posing a key challenge to avoiding interference with foreground on-demand transfers, as speculative traffic may contend for network and compute resources and reduce foreground decoding efficiency.

\begin{figure}[t]
    \centering
    \includegraphics[width=\columnwidth]{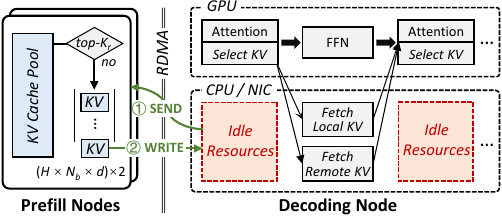}
    \caption{The speculative transfer utilizing idle resources.}
    \label{pic:speculative-transfer}
\end{figure}

% \textbf{\textit{Idle resources during attention computations.}}
Our key observation is that \textit{attention computation exposes idle network and CPU resources}, which can be leveraged to perform speculative transfers without impacting foreground execution.
The right part of Figure~\ref{pic:speculative-transfer} shows the operation flow of the decoding node, where InfiniGen's prefetch technique~\cite{osdi2024infinigen} is adopted.
Specifically, in each attention layer, the attention is executed concurrently with the generation of KV indexes for the next layer, \ie the KV selection process, using separate GPU streams.
The generated KV indexes are used to fetch the selected local and remote KV entries, as introduced in Section~\ref{sec:on-demand-transfer}, which overlaps with the FFN computation. 
Once all selected KV entries are available, the next attention layer starts, and the process repeats.
Since KV fetching can only begin after KV indexes are generated, the CPUs and RDMA NIC remain idle during the attention computation.

\textbf{\textit{Non-intrusive speculative KV transfer.}}
Based on the above observation, we propose transferring remaining KV entries from prefill nodes to the decoding node, utilizing idle resources.
At the start of each attention layer, the decoding node notifies all prefill nodes that the network is idle via RDMA \textsf{SEND} operations.
After \textsf{RECV}ing the notification, the prefill node starts transferring the remaining KV blocks that are not sent during the prefill stage.
Similar to the profile-based proactive transfer, these KV blocks are sent via RDMA \textsf{WRITE}s in the order of their importance, as defined in Equation~\ref{eq:importance}.
The corresponding KV masks are updated via additional \textsf{WRITE}s as introduced in Section~\ref{sec:proactive-transfer}.

% 介绍阈值
% To prevent speculative transfer from interfering with remote KV fetching, we limit each speculative transfer to only deliver a certain percentage of total KV blocks, which is defined as the \textit{speculative ratio}.
To prevent speculative transfer from interfering with remote KV fetching, it is crucial to terminate it in time.
Since notifying prefill nodes is not timely due to network round-trip time, we explicitly limit the number of KV blocks to send each time.
Specifically, each speculative transfer is restricted to send a fixed fraction of total KV blocks, defined as the \textit{speculative ratio}.
A lower speculative ratio mitigates interference, but requires more iterations to fully transfer the KV cache.
We set the ratio to 10\% to deliver all remaining KV entries to the decoding node within 10 decoding iterations, with almost no interference with on-demand KV fetching.

%%%%%%%%%%%%%%%%%%%%%%%%%%%%%%%%%%%%%%%%%%%%%%%%%%%%%%%%%%%
\subsection{Discussions}\label{sec:discussion}

\noindent
% \textbf{\textit{Generality of the \SmartGen design.}}
\textbf{\textit{Integrating different KV selection algorithms.}}
% 可以适用于不同KV选择算法 (依靠统一的KV index抽象)
\SmartGen can generalize to various KV selection algorithms and LLMs.
Specifically, it is compatible with various dynamic KV selection algorithms~\cite{osdi2024infinigen, acl2025hata, arxiv25retroinfer, arxiv2024clusterkv} through the unified KV index abstraction shown in Figure~\ref{pic:background-kv-selection}.
Besides, its generality to different LLMs is inherited from the KV selection algorithm it adopts.
These algorithms generally require metadata tensors, \ie lightweight auxiliary tensors derived from model weights or KV caches, to approximate attention scores without accessing full KV entries~\cite{osdi2024infinigen, acl2025hata, arxiv25retroinfer}.
These tensors are used in the decoding stage of \SmartGen to guide KV selection.
We recompute or pre-load weight-related metadata on the decoding node to save network bandwidth.
The key-cache-related metadata is transferred during the prefill stage.
For example, using a partial ratio of 0.3 in InfiniGen~\cite{osdi2024infinigen} results in a 15\% increase in KV cache transfer overhead.
This overhead is jointly considered by the analytical model and optimized by the selective transfer design.

\noindent
% \textbf{\textit{Generality of the offline profiling.}}
% \textbf{\textit{Improving the stability of offline profiling.}}
\textbf{\textit{Adapting to various workloads.}}
% The idea of positional similarity used in offline profiling has been validated and applied by many static KV selection methods~\cite{iclr2024streamingllm, iclr25duoattn}.
% Many recent models~\cite{deepseek-nsa} also incorporate a static component for KV selection.
% These approaches have demonstrated the model and workload stability of static KV importance. 
% Two methods can be adopted to further improve the stability of offline profiling:
In real-world LLM inference, request workloads are dynamic. Our offline profiling maintains stability to some extent, as positional similarity has been validated by prior static KV pruning methods~\cite{iclr2024streamingllm, iclr25duoattn} and recent models~\cite{deepseek-nsa}.
To handle larger workload variations, \SmartGen adopts:
(1) \textit{Periodic profiling}. The offline profiling can be performed periodically to adapt to workload changes by updating the importance matrix $I$ shown in Figure~\ref{pic:proactive-transfer}.
This process is transparent to the \SmartGen design, as the matrix $I$ can be updated using a standard \textit{read-copy-update} scheme.
(2) \textit{Grouped profiling}. The calibration datasets can be partitioned into additional groups based on prompt length, generating multiple importance matrices accordingly.
When serving online requests, \SmartGen selects the importance matrix from the most similar group.

\noindent
% \textbf{\textit{Adaptability of \SmartGen to various loads.}}
\textbf{\textit{Adapting to various network loads.}}
% 在网络负载很低时可以fall back (依靠现有的analytical model方案)
In LLM inference systems, network loads typically vary dynamically with the number of user requests and the changes of P/D settings.
\SmartGen can adapt to dynamically changing network loads based on the analytical model used by existing systems~\cite{sosp2024loongserve, osdi2024distserve, fast2025mooncake}.
Under high network loads, \SmartGen reduces stage-transition stalls through selective KV cache transfer, while under low loads, it automatically falls back to full KV cache transfer according to Equation~\ref{eq:top-k}.
Moreover, by employing more advanced analytical models, \SmartGen could save precious network bandwidth for other important tasks, \eg prefix cache transfer~\cite{fast2025mooncake}. 
In contrast, existing quantization-based KV cache transfer schemes~\cite{sigcomm25hack, sigcomm2024cachegen} cannot adapt to dynamic network conditions, as they cannot adjust KV cache quantization precision in response to changing loads at runtime, nor easily compensate for accuracy loss via on-demand transfer.

% \noindent
% \textbf{\textit{Adaptability of \SmartGen to various parallel strategies.}}
% As a transfer engine, \SmartGen only needs the source and destination addresses for KV cache transfer.
% By knowing how KV cache blocks are split and concatenated under different parallelism schemes, \SmartGen can correctly determine the source and destination of each block.
% For DP, KV caches are concatenated along the batch dimension (rows).
% For tensor parallelism (TP), KV caches are concatenated along the model dimension (columns).
% For pipeline parallelism (PP), no concatenation is needed, as each prefill instance holds a KV cache from a different layer.

% \noindent
% \textbf{\textit{Model accuracy in \SmartGen.}}
% % 并没有改变推理精度，能保持和所选KV选择算法完全相同的输出
% % 最终的完全传输甚至带来了提高模型精度的可能性 (由于与本工作正交，我们把它留给future works)
% % Thanks to the on-demand transfer during the decoding stage, \SmartGen does not compromise inference accuracy.
% % It produces exactly the same generation outputs as the adopted KV selection algorithm, \eg InfiniGen~\cite{osdi2024infinigen}.
% \SmartGen maintains the same accuracy as InfiniGen~\cite{osdi2024infinigen} and any other adopted KV cache selection algorithms.
% Additionally, since the full KV cache is finally speculatively delivered to the decoding node, \SmartGen could even increase the number of selected tokens after receiving the full KV cache to improve accuracy.
% We leave this as future work, as it is orthogonal to the \SmartGen design.

\noindent

\noindent
\textbf{\textit{Supporting cloud instances lacking GDR.}}
Some low-cost cloud servers do not support GDR due to virtualization constraints.
In such environments, the parallel on-demand transfer design requires two modifications:
% , as shown in Figure~\ref{pic:on-demand-transfer-wo-gdr}:
(1) \textit{On-CPU index splitting.}
Since prefill nodes cannot update the KV mask matrix via GDR, the KV mask should be maintained in host memory. Consequently, the KV index is offloaded to host memory before being split.
(2) \textit{Extra local KV loading.}
Since the selected remote KV entries cannot be directly transferred into the GPU, the decoding node should first receive them in host memory and then load them into GPU via an extra host-to-device copy.
Given that PCIe bandwidth is much higher than the network bandwidth, these overheads are acceptable.
\section{Evaluation}\label{sec:evaluation}

\begin{figure*}[t]
    \vspace{3mm}
    \centering
    \includegraphics[width=1.0\textwidth]{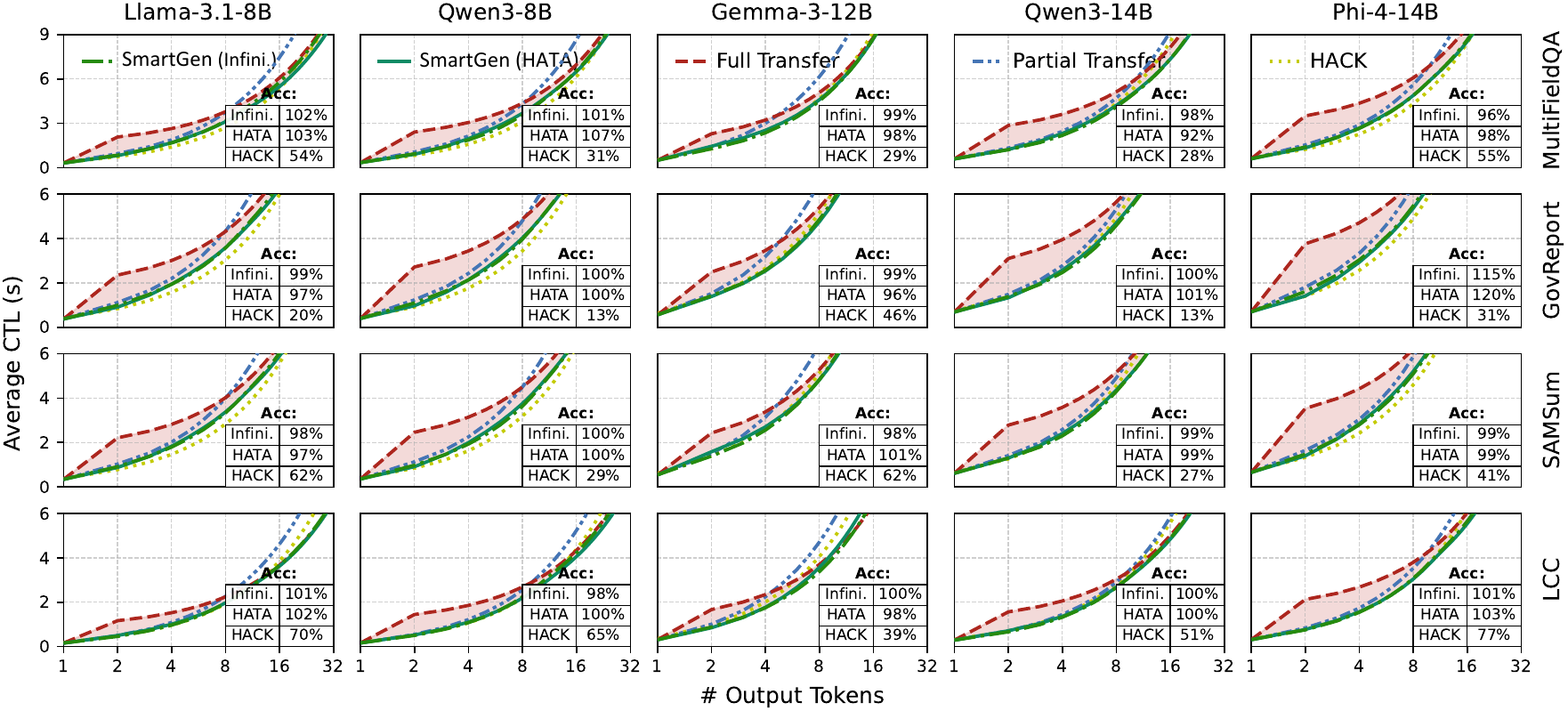}
    \caption{The average request CTL over time during token generation. The bottom-right table shows the accuracy relative to the full-cache baseline. More accuracy results will be discussed in \S~\ref{sec:accuracy}.}
    \label{fig:ctl}
    \vspace{-5mm}
\end{figure*}

% \begin{figure*}[t]
%     \centering
%     \includegraphics[width=1.0\textwidth]{Figures/ctl_acc.pdf}
%     \caption{The average request CTL during token generation (upper half) and the accuracy (lower half) on the LongBench benchmark~\cite{acl2024longbench}.
%     The red shading shows the stage-transition stall, and the red numbers indicate the TTST improvement of \SmartGen (HATA) over full transfer.
%     }
%     \label{fig:ctl-acc}
% \end{figure*}

\subsection{Experimental Setup}\label{sec:exp-setup}

\noindent
\textbf{\textit{Testbed.}}
% We conduct all experiments on 4 physical machines (\ie 3 prefill nodes and 1 decoding node) on CloudLab~\cite{atc2019cloudlab}.
% Each machine is equipped with two NVIDIA Tesla V100S GPUs (32 GB memory each), two 32-core AMD EPYC 7542 CPUs, 512 GB of DRAM, and a 25 Gbps Mellanox ConnectX-5 NIC.
% Within each machine, the CPUs, GPUs, and NIC are interconnected via PCIe 3.0 $\times16$.
% All machines are connected through a 25 Gbps Ethernet switch.
% Such a setting aligns with the inter-node network bandwidths in major cloud providers.
The Alibaba GPU instances~\cite{alibaba-cloud-instances} used in this paper are listed in Table~\ref{tab:GPU-instances}.
Unless otherwise stated, we conduct experiments on 3 \textit{ecs.gn8is-2x.8xlarge} prefill instances and one \textit{ecs.gn8is.4xlarge} decoding instance.
% Each prefill instance is equipped with two NVIDIA L20 GPUs (48 GB memory each), 32 vCPUs, 256 GB of DRAM, and an eRDMA interface~\cite{Alibaba-eRDMA} providing 32 Gbps network bandwidth.
Each instance is equipped with an eRDMA interface~\cite{Alibaba-eRDMA}.
Within each instance, the vCPUs, GPUs, and vNIC are interconnected via PCIe 4.0$\times$16.
% If more powerful GPUs are adopted under the same network settings, the speedup in computation can only make the network bottleneck worse (as simulated in Figure~\ref{fig:network-congestion}), and \SmartGen could achieve better performance.
For each prefill node, we launch two processes with each on one GPU, achieving a maximum DP degree of 6 in total.
For the decoding node, we adopt one GPU with larger batch sizes to maximize the GPU utilization~\cite{osdi2024distserve, isca2024splitwise, arxiv2024tetriinfer}.

\noindent
\textbf{\textit{Models and workloads.}}
% Similar to prior works on LLM inference~\cite{osdi2024infinigen, fast2025impress, osdi2024distserve, eurosys2025hcache}, we use OPT~\cite{arxiv2022opt} and Llama-2~\cite{arxiv2023llama-2} models for evaluation, which are representative LLM families widely used in academia and industry.
% We chose these two model series since InfiniGen~\cite{osdi2024infinigen} has only open-sourced its adaptation details for these models.
% Using more cache-efficient models could complement \SmartGen in alleviating the network bottleneck, enabling \SmartGen to \textit{scale} to larger models with longer sequence lengths\footnote{Like prior works~\cite{osdi2024distserve}, we truncate overlong prompts because OPT models only support a maximum sequence length of 2K tokens.}
% and larger batch sizes.
% \SmartGen's design remains effective for more KV cache-efficient models, which face similar network bottlenecks with longer sequences~\cite{llama-4, qwen-2.5-turbo, megabean-mistral},\footnote{Like prior works~\cite{osdi2024distserve}, we truncate overlong prompts because OPT models only support a maximum sequence length of 2K tokens.}
% larger batches, and larger model sizes.
We use Qwen3~\cite{qwen3}, Meta's Llama-3.1~\cite{llama-3}, Google's Gemma-3~\cite{gemma-3}, and Microsoft's Phi-4~\cite{phi-4} models for evaluation, which are representative LLM families widely used in academia and industry.
As for workloads, we use the LongBench benchmark~\cite{acl2024longbench}, which covers a wide range of long-context tasks.
We select the alphabetically first dataset from each of the four tasks, \ie \textit{MultiFieldQA} (document query answering), \textit{GovReport}~\cite{govreport} (summarization), \textit{SAMSum}~\cite{samsum} (few-shot learning), and \textit{LCC}~\cite{lcc} (code completion), to comprehensively assess \SmartGen's performance across diverse scenarios.
The average prompt lengths for the four workloads are 7K, 10K, 9K, and 3K tokens, respectively, with up to 60K tokens batched on our testbed by default.
Besides, we separate LongBench's first subset (\ie \textit{2WikiMultihopQA}~\cite{2wikimqa}) as a dedicated calibration dataset in advance to ensure it is not used during evaluation.

\begin{table}[t]
    \centering
    \caption{GPU instances on the cloud used in this paper.}
    \resizebox{\linewidth}{!}{
        \begin{tabular}{@{}ccccc@{}}
        \toprule
        \textbf{Name}                 & \textbf{GPUs} & \textbf{vCPU} & \textbf{DRAM} & \textbf{Network}  \\ \midrule
        \textit{ecs.gn8is-2x.8xlarge} & 2 L20 (2*48 GB) & 32            & 256 GB        & 32 Gbps          \\ \midrule
        \textit{ecs.gn8is.4xlarge}    & 1 L20 (48 GB) & 16            & 128 GB        & 25 Gbps           \\ \midrule
        \textit{ecs.gn8is.2xlarge}    & 1 L20 (48 GB) & 8             & 64 GB         & 15 Gbps           \\ \bottomrule
        \end{tabular}
    }
    \label{tab:GPU-instances}
\end{table}

% For performance evaluation, we randomly sample requests from three real-world workloads: LongBench~\cite{acl2024longbench}, L-Eval~\cite{acl2024leval}, and ShareGPT~\cite{sharegpt}.
% Like prior work~\cite{sosp2024loongserve, fast2025mooncake}, we also construct a mixed workload by uniformly sampling from the three.
% The average sampled prompt lengths for the four workloads are 1903, 2005, 768, and 1577 tokens, respectively.
% Besides, we separate LongBench's first subset as a dedicated calibration dataset in advance to ensure it is not used during evaluation.

% For accuracy evaluation, we use the LongBench benchmark~\cite{acl2024longbench}, which targets various long-context tasks.
% We select the alphabetically first dataset from each of the four tasks, \ie \textit{MultiFieldQA} (document query answering), \textit{GovReport}~\cite{govreport} (summarization), \textit{SAMSum}~\cite{samsum} (few-shot learning), and \textit{LCC}~\cite{lcc} (code completion), to comprehensively assess \SmartGen's accuracy across diverse scenarios.

\noindent
\textbf{\textit{Comparisons.}}
% We compare \SmartGen with three baselines:
We compare the following four schemes:
\begin{itemize}[noitemsep, topsep=0pt, parsep=0pt, partopsep=0pt]
    \item \textbf{\SmartGen (Infini./HATA)}: We implement \SmartGen by adopting InfiniGen~\cite{osdi2024infinigen} and HATA~\cite{acl2025hata}, respectively, to verify the generality of the \SmartGen design with either training-free or training-based KV selection algorithms.
    \item \textbf{Full transfer}: This is the typical KV cache transfer scheme that pushes all KV cache layer-by-layer during the prefill stage. To our knowledge, this is the state-of-the-art scheme widely adopted in existing disaggregated LLM inference systems~\cite{fast2025mooncake, nvidia-dynamo, isca2024splitwise, icml2024dejavu, arxiv2024tetriinfer}.
    \item \textbf{Partial transfer}: This is the vanilla selective KV cache transfer scheme that pushes the first sequential $K_r$ KV blocks (\ie in memory address order) during prefill, and fetches the missing entries on demand during decoding.
    \item \textbf{HACK}~\cite{sigcomm25hack}: This is the state-of-the-art quantization-based scheme that addresses the network bottleneck in disaggregated inference for self-hosted LLMs on the cloud. It quantizes the KV cache to \textit{int2} and the query to \textit{int8} using homomorphic quantization, enabling it to store the 2-bit KV cache on the GPU without offloading.
\end{itemize}
For fairness, all methods are implemented on top of the same LLM inference system~\cite{icml2023flexgen}.
FlashAttention~\cite{nips2022flashattention} and FlashInfer~\cite{arxiv25flashinfer} are adopted to improve inference performance.
On the prefill node, we assume a prefix cache hit ratio of 75\% on host memory to improve prefill performance~\cite{arxiv2025cachewild, fast2025mooncake, nips2024sglang, atc2024cachedattention, eurosys25pensieve}.
The KV caches to be transferred are partitioned using the same block granularity.
On the decoding node, we improve the computational efficiency of full transfer by applying local KV cache selection using InfiniGen~\cite{osdi2024infinigen}.
% all methods adopt the same KV selection algorithm~\cite{osdi2024infinigen} to improve decoding efficiency.

% As for accuracy, we compare \SmartGen with the following three additional attention schemes:
% \begin{itemize}[noitemsep, topsep=0pt, parsep=0pt, partopsep=0pt]
%     \item \textbf{Full Cache}: This is the full attention computation without adopting any sparse attention algorithms or quantizations. It represents the optimal accuracy.
%     \item \textbf{StreamingLLM}~\cite{iclr2024streamingllm}: This is a classical static KV sparsity algorithm. It keeps only the initial tokens and a sliding window of recent tokens for attention. For simplicity, one KV block is kept at the beginning.
%     \item \textbf{Proactive-only (Infini./HATA)}: This baseline simply evicts KV cache entries unselected by the offline profiling and compute attention without on-demand compensation.
% \end{itemize}

\noindent
\textbf{\textit{Key metrics.}}
We evaluate the average \textit{cumulative per-token latency} (CTL) per request to assess \SmartGen's overall performance for users.
We focus on addressing the stage-transition stall issue to achieve a seamless disaggregated LLM inference.
Therefore, we also evaluate the \textit{time-to-second-token} (TTST) to show the performance of the KV cache transfer, and the \textit{time-between-tokens} (TBT) to show the performance overhead induced by \SmartGen.
To isolate the overhead, we exclude TTST during the TBT calculation.
As for accuracy, we use LongBench's accuracy-related metrics (\%) to measure the impact of KV selection in \SmartGen across different datasets.

\noindent
\textbf{\textit{Parameters.}}
We use the suggested configurations of InfiniGen and HATA, \eg an alpha value of 5, a partial ratio of 0.3, a hash bit count of 256, and a maximum KV selection ratio of 20\%.
Unless otherwise specified, we set the number of KV blocks per layer (\ie $M$) to 1K and adopt a DP degree of 6.
All requests are configured with an output length of 64 tokens.
As for \SmartGen, we set the speculative ratio to 10\%.

\begin{figure*}[t]
    \hspace{1.5mm}
    \begin{minipage}[b]{0.23\textwidth}
        \centering
        \includegraphics[width=\textwidth]{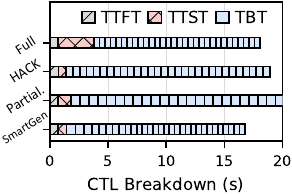}
        \caption{The CTL breakdown.}
        \label{fig:latency-breakdown-test}
    \end{minipage}
    \hspace{-1mm}
    \begin{minipage}[b]{0.405\textwidth}
        \centering
        \includegraphics[width=0.5\textwidth]{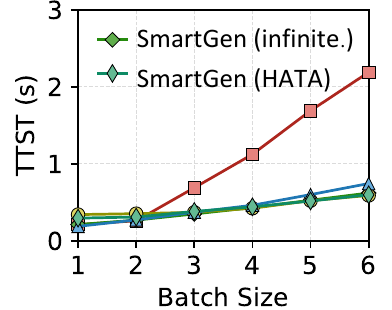}
        \hspace{-2mm}
        \includegraphics[width=0.5\textwidth]{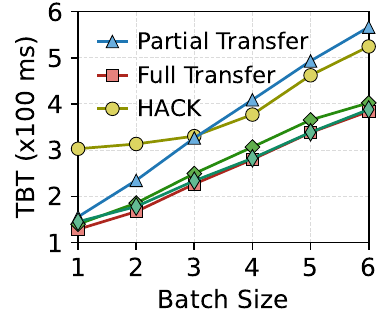}
        \caption{Performances with various batch sizes.}
        \label{fig:batch-size-test}
    \end{minipage}
    \hspace{-2mm}
    \begin{minipage}[b]{0.36\textwidth}
        \centering
        \includegraphics[width=0.9\textwidth]{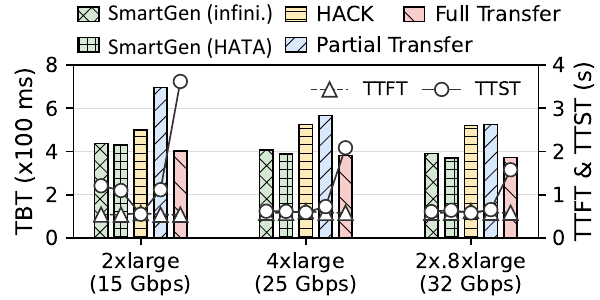}
        \caption{Performances with various bandwidths.}
        \label{fig:various-bandwidth-test}
    \end{minipage}
\end{figure*}

\begin{figure*}[t]
    \vspace{-6.5mm}
    \begin{minipage}[b]{0.75\textwidth}
        \centering
        \subfloat[L20 instances (25 Gbps virtualized network).
        % (25 Gbps network w/o GDR).
        ]{
            \includegraphics[width=0.49\textwidth]{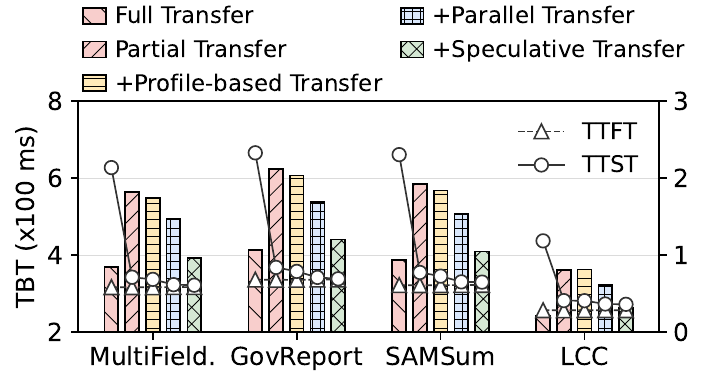}
            \label{fig:factor-analysis-L20}
        }
        \hspace{-4.5mm}
        \subfloat[V100S instances (25 Gbps physical network).
        % (25 Gbps network w/ GDR).
        ]{
            \includegraphics[width=0.497\textwidth]{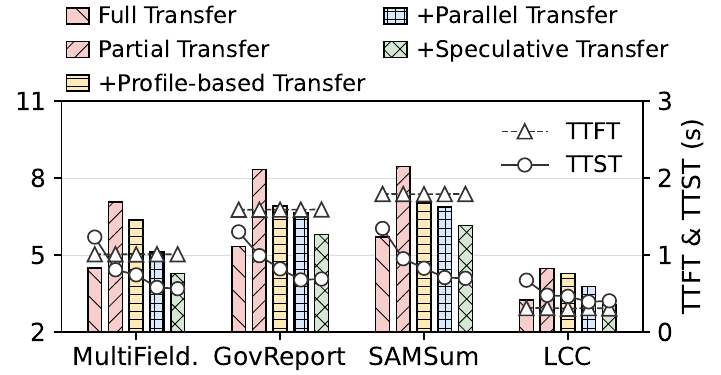}
            \label{fig:factor-analysis-V100}
        }
        \caption{The factor analysis for techniques in \SmartGen on various GPU instances.}
        \label{fig:factor-analysis}
    \end{minipage}
    \hspace{-4.4mm}
    \begin{minipage}[b]{0.265\textwidth}
        \centering
        \includegraphics[width=0.95\textwidth]{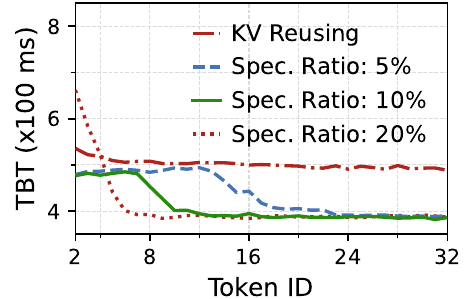}
        \caption{The effectiveness of speculative transfer.}
        \label{fig:speculative-effectiveness}
    \end{minipage}
    \vspace{-8.5mm}
\end{figure*}

\begin{figure}[t]
    \centering
    \includegraphics[width=0.95\columnwidth]{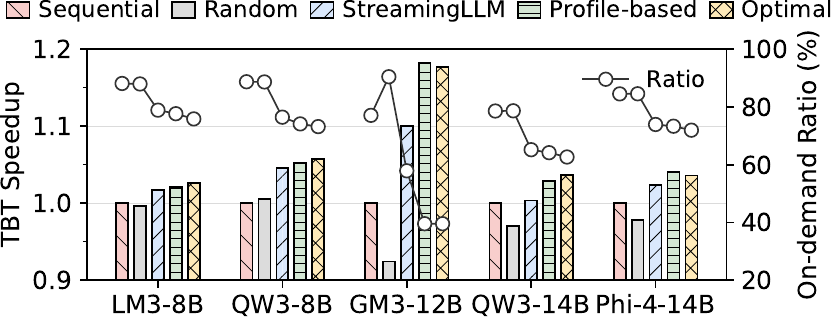}
    \caption{The effectiveness of profile-based proactive transfer.}
    \label{fig:profiling-effectiveness}
\end{figure}

\begin{figure*}[t]
    \centering
    \subfloat[Sequence length.]{
        \includegraphics[width=0.170\textwidth]{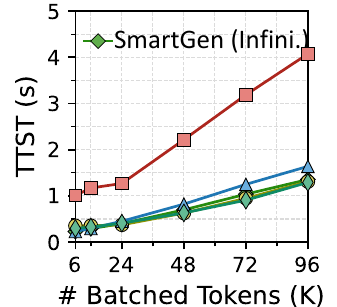}
        \hspace{-2.9mm}
        \includegraphics[width=0.176\textwidth]{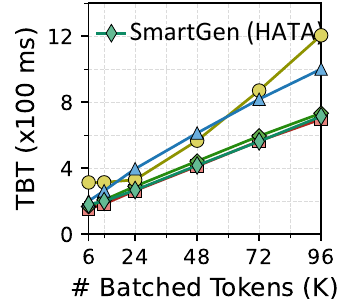}
        \label{fig:impact-of-seq-len}
    }
    \hspace{-3.7mm}
    \subfloat[Number of KV blocks.]{
        \includegraphics[width=0.166\textwidth]{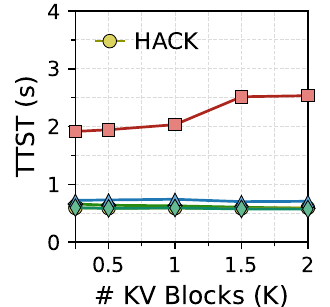}
        \hspace{-3mm}
        \includegraphics[width=0.166\textwidth]{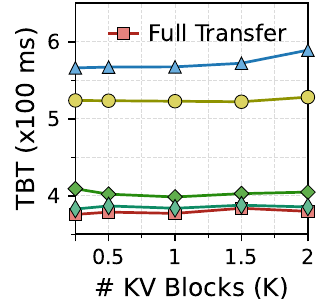}
        \label{fig:impact-of-block-num}
    }
    \hspace{-3.85mm}
    \subfloat[KV selection ratio.]{
        \includegraphics[width=0.169\textwidth]{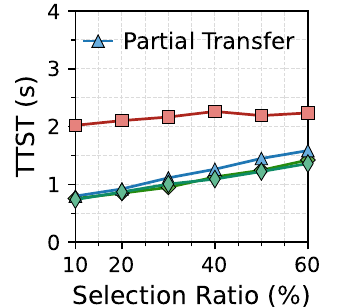}
        \hspace{-2.6mm}
        \includegraphics[width=0.175\textwidth]{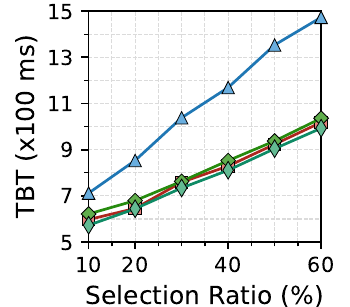}
        \label{fig:impact-of-alpha-value}
    }
    \caption{The sensitivity analysis for overall performance.}
    \label{fig:sentivity-analyses}
    \vspace{-5mm}
\end{figure*}

%%%%%%%%%%%%%%%%%%%%%%%%%%%%%%%%%%%%%%%%%%%%%%%%%%%%%%%%%%%%%%%%%%%%%%%%%%%%%%%%%%%%%%%%%%%%%
\subsection{Performance Comparison}\label{sec:performance-comparison}
\noindent
% Figure~\ref{fig:ctl} shows the performance of all schemes across different models and workloads with a batch size of 6.
Figure~\ref{fig:ctl} shows the performance of all schemes across different models and workloads on the LongBench workloads.
Figures~\ref{fig:latency-breakdown-test}-\ref{fig:various-bandwidth-test} provide detailed analyses of TTST and TBT under the MultiFieldQA workload on the Qwen3-14B model.
Similar trends are observed on other models and workloads.

% Figure~\ref{fig:batch-size-test} presents the performance of all transfer schemes on the OPT-13B model across the four workloads, where we set the DP degree equal to the batch size to ensure stable TTFT.
% The performance of \SmartGen and other baselines follows the same trend on all workloads.
% Without loss of generality, we focus our discussion only on L-Eval and ShareGPT.

\textbf{\textit{CTL results.}}
% 整体解读
The average request CTL over time reflects user-perceived performance, as shown in Figure~\ref{fig:ctl}.
% To better illustrate the impact of CTL on user experience, we use a curve representing a constant reading speed of 4 tokens per second to simulate human reading~\cite{jmr2019many}, as shown in Figure~\ref{fig:ctl}.
% A CTL curve that closely follows or stays below the human-reading curve indicates smooth token generation.
Full transfer exhibits an early CTL spike, suggesting users experience stalls (shown as shadow) when transferring the entire KV cache.
\SmartGen reduces TTST by up to $4.3\times$ on GovReport compared with full transfer, as shown in Figure~\ref{fig:latency-breakdown-test}, indicating a seamless inference.
Partial transfer shows a generally higher CTL due to the overhead of on-demand transfers.
% HACK aggressively quantizes the KV cache to reduce the transfer bottleneck, and thus it achieves only up to 77\% relative accuracy on the LongBench benchmark, which is significantly lower than InfiniGen and HATA.
HACK alleviates the transfer bottleneck via aggressive KV cache quantization, achieving substantial efficiency gains, but at the cost of some accuracy, reaching up to 77\% relative accuracy on LongBench, lower than InfiniGen and HATA.
This is likely because LongBench requires precise capture of key information in long contexts and is thus more sensitive to quantization errors.
Besides, although HACK avoids offloading, it requires unpacking the KV cache from a compact format to a computation-ready format at each step, resulting in a higher TBT than full transfer and even \SmartGen in some cases.
In contrast, \SmartGen progressively reduces TBTs and keeps the CTL generally the best, thanks to optimized KV cache transfers at each stage.

\textbf{\textit{TTST results.}}
% 先方法间比较
The left half of Figure~\ref{fig:batch-size-test} shows the TTST results under the MultifieldQA workload on Qwen3-14B. 
The DP degree is set equal to the batch size to maintain stable TTFT.
With a batch size of 6, \SmartGen, HACK, and partial transfer achieve $3.7\times$, $3.7\times$, and $2.9\times$ lower TTST, compared with full transfer.
This is because they do not transfer the entire KV cache from prefill nodes to the decoding node, saving network bandwidth.
Besides, \SmartGen outperforms partial transfer by $1.2\times$ due to the more efficient on-demand transfer when decoding the second token.
% 再趋势
As the batch size grows from 1 to 6, TTSTs of full transfer, partial transfer, HACK, and \SmartGen increase by $10.4\times$, $4.0\times$, $1.7\times$, and $2.0\times$, respectively.
The increase in TTST for full transfer is attributed to the more KV caches to transfer as the batch size grows.
In contrast, increases observed in \SmartGen, HACK, and partial transfer are mainly due to the higher memory access intensity on the decoding node.

\textbf{\textit{TBT results.}}
The right half of Figure~\ref{fig:batch-size-test} shows the average TBTs under the MultifieldQA workload on Qwen3-14B.
% 先方法间比较
The TBT of full transfer represents the ideal case since all KV caches are locally available on the decoding node.
With a batch size of 6, partial transfer exhibits a $1.5\times$ higher TBT than \SmartGen and full transfer, respectively, as it requires additional network round-trips on the critical path of every decoding iteration.
HACK incurs a $1.4\times$ higher TBT than both \SmartGen and full transfer due to format conversion overhead.
% , which scales with the size of the KV cache that requires conversion.
\SmartGen achieves a TBT close to the ideal case thanks to its effective and efficient designs on proactive, on-demand, and speculative transfers.
% Similar to the TTST results, Figure~\ref{fig:opt-13b-tbt-sharegpt} shows that all schemes exhibit similar TBTs on ShareGPT. %, with performance differences within $1.2\times$.
% This is because shorter prompt lengths require fewer KV entries to be fetched from prefill nodes on demand, allowing the limited network bandwidth to handle the transfer efficiently.
% 再趋势
As the batch size grows from 1 to 6, TBTs of full transfer, partial transfer, HACK, and \SmartGen increase by $3.0\times$, $3.6\times$, $1.7\times$, and $2.7\times$, respectively, due to increased memory access intensity.

\textbf{\textit{Performance with various network bandwidths.}}
Figure~\ref{fig:various-bandwidth-test} shows the performance of all schemes across decoding instances with different network bandwidths, verifying \SmartGen's ability to adapt to varying network conditions.
As the network bandwidth decreases from 32 Gbps to 15 Gbps, \SmartGen's TTST improvement over full transfer increases from $2.5\times$ to $3.3\times$, and the TBT improvement over partial transfer increases from $1.4\times$ to $1.6\times$.
At 15 Gbps, the TTST of \SmartGen exceeds that of the ideal case.
This is because the bandwidth is saturated by the transfer of the first two layers of KV cache and the metadata tensors required by the KV selection algorithms. We leave the optimization of this overhead for future work.

\subsection{Factor Analysis}\label{sec:factor-anlysis}
\noindent
Figure~\ref{fig:factor-analysis} presents the factor analysis for \SmartGen on Qwen3-14B.
% Since the L20 instances lack GDR support, 
To demonstrate \SmartGen's performance across different hardware platforms, this section additionally evaluates it on V100S physical servers, \ie \textit{r7525} instances on CloudLab~\cite{atc2019cloudlab}, using the same amount of GPU memory for prefilling.
Each proposed technique is applied to the partial transfer one by one.
% We also show the performance of full transfer as a baseline.
% Due to the space limit, we only analyze results on the L-Eval workload.
% Results on other workloads exhibit similar trends.
We analyze results only on MultifieldQA due to space limits.
Others exhibit similar trends.

\textbf{\textit{Partial transfer.}}
Compared with full transfer, partial transfer reduces TTST at the cost of increased TBT.
It achieves a $3.0\times$ and $1.5\times$ reduction in TTST but incurs a $1.5\times$ and $1.6\times$ increase in TBT on L20 and V100S instances, respectively.
This trade-off arises since fewer KV entries are transferred during prefill, necessitating transfers of missing entries during the decoding stage.
% The TBT penalty on the Llama-2-13B model is higher than that on the OPT-13B model, as InfiniGen identifies a larger number of important KV entries on Llama models.
% The TBT penalty on V100S instances is higher than that on L20 since V100S instances offer lower network bandwidth.
The following designs address the TBT penalty introduced by partial transfer.

\textbf{\textit{+ Profile-based proactive transfer.}}
The profile-based proactive transfer reduces TBT by $1.03\times$ and $1.1\times$ on L20 and V100S instances, respectively, by prioritizing the proactive transfer of important KV entries.
The TBT improvements on L20 instances are lower than those on V100S instances.
This is likely because elastic NICs on Alibaba Cloud offer a less stable bandwidth ceiling than physical NICs on CloudLab, leading to smaller performance gains.

Figure~\ref{fig:profiling-effectiveness} compares different KV selection strategies for prefill nodes across various models under the MultiFieldQA workload on L20 instances.
% On Llama-2 (LM) models, the sequential strategy performs even worse than the random one, as it fails to select KV caches from the last several attention layers.
% However, these layers also contain many important KV entries, as verified in Figure~\ref{fig:profiling}.
The profile-based strategy achieves up to $1.3\times$ and $1.2\times$ speedups over the random and sequential strategies, respectively, by reducing the on-demand ratio by up to 51\% and 38\%.
StreamingLLM~\cite{iclr2024streamingllm}, a static KV sparsity strategy, yields large gains on Gemma-3-12B due to its alignment with the model's sliding-window attention components.
In addition, the profile-based strategy achieves up to a $1.1\times$ speedup over StreamingLLM and reduces the on-demand ratio by up to 18\% by further identifying important KV entries in full attention.
It also achieves performance close to the optimal case, validating the effectiveness of profiling positional similarity.

\textbf{\textit{+ Parallel on-demand transfer.}}
The parallel on-demand transfer reduces TBT by $1.1\times$ and $1.2\times$ on L20 and V100S instances, respectively.
This is because it removes the network round-trip for remote KV fetching from the critical path of local KV cache loading during decoding.
% The improvement on V100S is more pronounced for two reasons:
% 1) Enabling GDR accelerates the execution of remote lanes without requiring the launch of an additional kernel.
% 2) The lower PCIe bandwidth (\ie PCIe 3.0$\times$16) on V100S instances prolongs the duration of local lanes, allowing \SmartGen to better hide the parallel network round-trip for remote KV fetching.

\textbf{\textit{+ Speculative Transfer.}}
The speculative transfer further brings $1.3\times$ and $1.2\times$ reduction in TBT on L20 and V100S instances, respectively, by speculatively delivering all KV entries to the decoding node.
Figure~\ref{fig:speculative-effectiveness} shows the details on Qwen3-14B under the MultifieldQA workload.
Unlike KV reusing, \ie reusing previously fetched KV entries in host memory, speculative transfer proactively delivers all remaining KV blocks within a few iterations, after which the TBT aligns with the ideal case.

With a speculative ratio of 20\%, speculative transfer completes within 5 iterations.
However, TBTs of the first few tokens increase by up to $1.4\times$ since excessive speculative KV block transfers interfere with subsequent KV fetching.
In contrast, a speculative ratio of 5\% introduces no such interference but requires approximately 20 iterations to complete all transfers.
\SmartGen chooses an appropriate ratio of 10\%, enabling transferring all remaining KV blocks as early as possible (\ie 10 iterations) with minimal impact on TBTs.

%%%%%%%%%%%%%%%%%%%%%%%%%%%%%%%%%%%%%%%%%%%%%%%%%%%%%%%%%%%%%%%%%%%%%%%%%%%%%%%%%%%%%%%%%%%%%
\subsection{Sensitivity}
\noindent
This section investigates how some parameters affect the performance on Qwen3-14B and MultiFieldQA.

\textbf{\textit{Impact of sequence length.}}
% Figure~\ref{fig:impact-of-seq-len} shows the impact of sequence lengths.
% We control the sequence length by truncating overlong requests in the mixed workload.
Figure~\ref{fig:impact-of-seq-len} shows that \SmartGen consistently performs best with various sequence lengths.
As the number of batched tokens increases from 6K to 96K tokens, the TTST of full transfer grows more rapidly (3.1 s) compared with partial transfer, HACK, and \SmartGen (1.4/1.0/1.0 s), respectively, since it must transfer the entire KV cache, whose size scales linearly with the sequence length.
Meanwhile, the TBT of partial transfer and HACK increases more rapidly (0.8/0.9 s) than those of full transfer and \SmartGen (0.5 s).
This is because HACK needs to unpack all KV cache entries, and partial transfer fetches remote KV entries inefficiently, the overhead of which also increases with the sequence length.

% \SmartGen consistently performs best across varying sequence lengths by combining the TTST advantage of partial transfer with a TBT comparable to that of full transfer.

\textbf{\textit{Impact of number of KV blocks.}}
Figure~\ref{fig:impact-of-block-num} illustrates the impact of the number of KV blocks per attention layer (\ie $M$) on system performance.
% As $M$ decreases from 75 to 20, the TTST of full transfer increases by $3.0\times$, due to more severe network congestion under limited bandwidth.
% This also reduces the accuracy of the analytical model in predicting $T_t$, leading to a $2.0\times$ and $2.3\times$ increase in TTSTs of partial transfer and \SmartGen, respectively. 
The TTST of full transfer increases by $1.2\times$ when $M$ exceeds 1K, due to NIC processing overhead.
The performances of other schemes remain stable across different $M$ values.
This is because important KV entries within each layer tend to be spatially clustered, as shown in Figure~\ref{fig:profiling}, reducing the need for fine-grained profiling.
We set $M$ to 1K to mitigate its impact on full transfer.

\textbf{\textit{Impact of KV selection ratio.}}
As shown in Figure~\ref{fig:impact-of-alpha-value}, the TTST of full transfer remains consistently high across varying selection ratios, as it transfers all KV entries regardless of the KV selection algorithm, making the KV cache transfer overhead dominate the overall TTST.
As the ratio increases from 10\% to 60\%, TTSTs of partial transfer and \SmartGen increase by $2.0\times$ and $1.8\times$, and their TBTs grow by $2.1\times$ and $1.7\times$, respectively.
This is because higher selection ratios result in more KV entries being selected, leading to increased TBT, which in turn constitutes a larger proportion of the TTST.
\SmartGen maintains superior performance across varying sparsity degrees with both InfiniGen and HATA, implying its adaptability to more KV selection algorithms with diverse selection intensities.
% The alpha value is a hyperparameter in InfiniGen used to control the amount of important KV entries.
% For each head, tokens with attention scores exceeding the maximum score minus the alpha are selected~\cite{osdi2024infinigen}.
% The larger the alpha value, the more KV entries will be selected.
% As shown in Figure~\ref{fig:impact-of-alpha-value}, the TTST of full transfer remains consistently high across varying alpha values, as it transfers all KV entries regardless of alpha, making the KV cache transfer overhead dominate the overall TTST.
% As the alpha value increases from 1 to 9, TTSTs of partial transfer and \SmartGen increase by $4.8\times$ and $2.9\times$, and their TBTs grow by $5.8\times$ and $5.0\times$, respectively.
% This is because higher alpha values result in more KV entries being selected, leading to increased TBT, which in turn constitutes a larger proportion of the TTST.
% \SmartGen maintains superior performance across varying alpha values, implying its adaptability to KV selection algorithms with diverse selection intensities.

%%%%%%%%%%%%%%%%%%%%%%%%%%%%%%%%%%%%%%%%%%%%%%%%%%%%%%%%%%%%%%%%%%%%%%%%%%%%%%%%%%%%%%%%%%%%%

\begin{figure*}[t]
    \vspace{3mm}
    \centering
    \includegraphics[width=1.0\textwidth]{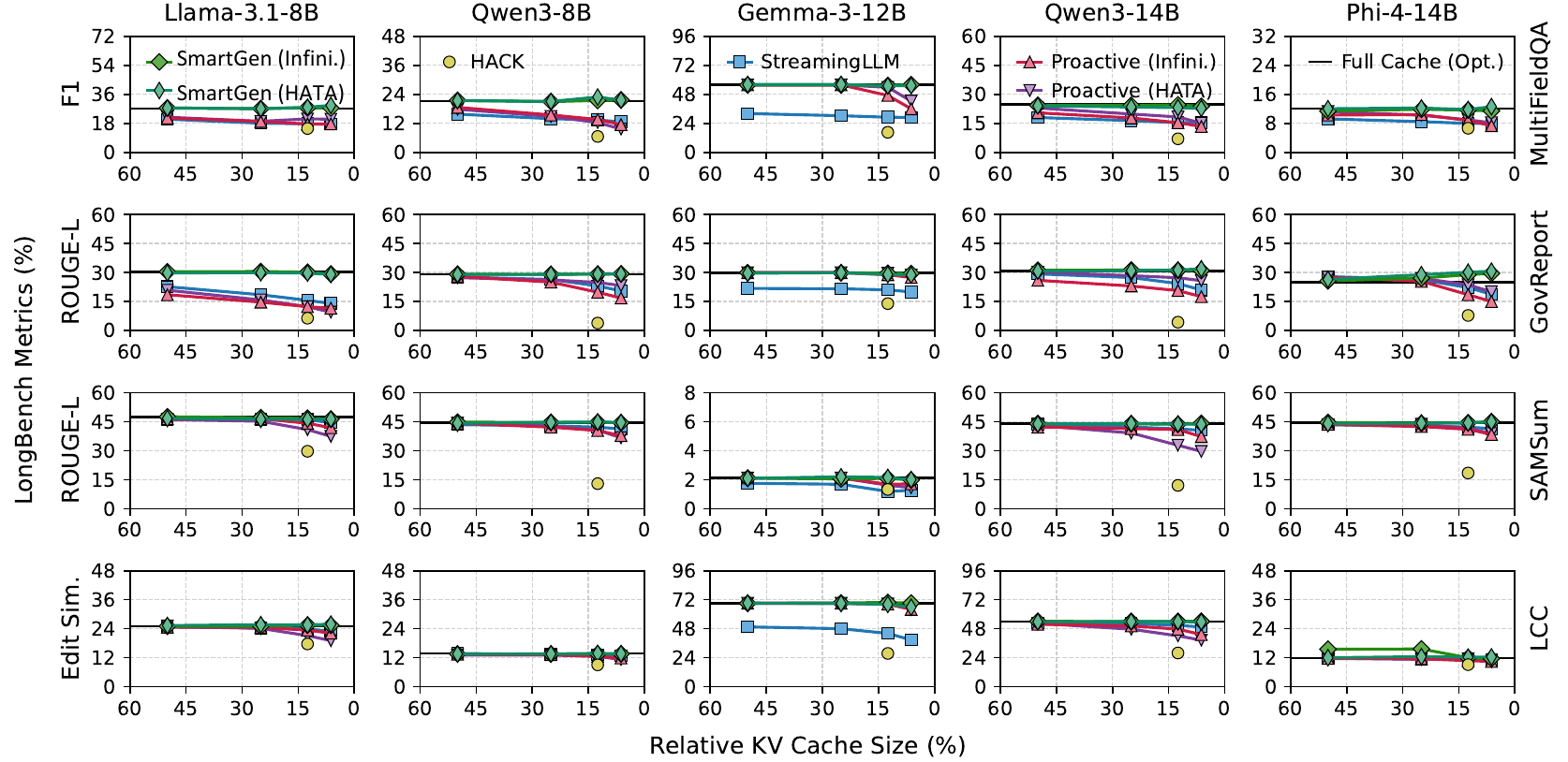}
    \caption{The accuracy analysis of LLMs on the LongBench benchmark~\cite{acl2024longbench}.}
    \label{fig:accuracy}
    \vspace{-5mm}
\end{figure*}

\subsection{Accuracy}\label{sec:accuracy}
\noindent
Figure~\ref{fig:accuracy} presents the accuracy of \SmartGen and some baselines across various models on LongBench.
The relative KV cache size indicates the ratio between the KV cache used in attention and that of the full-cache baseline.

\textbf{\textit{Overall accuracy.}}
\SmartGen consistently shows high accuracy across the models and tasks.
Its accuracy closely matches the full-cache baseline.
This is because \SmartGen directly adopts the state-of-the-art KV selection method, \ie InfiniGen~\cite{osdi2024infinigen} and HATA~\cite{acl2025hata}, without modifying its algorithm design.
In contrast, HACK shows lower accuracy with 2-bit quantization, demonstrating that dynamic KV selection is more effective than quantization in mitigating the KV cache transfer issue for challenging long-context understanding tasks where precision is critical.

\textbf{\textit{Profiling accuracy.}}
To validate the feasibility of offline profiling, we also assess two proactive-only baselines, \ie proactive (Infini./HATA), that simply evict KV cache entries unselected by the profiling.
They generally achieve accuracy comparable to StreamingLLM~\cite{iclr2024streamingllm}.
This is because offline profiling can accurately identify universally important tokens like static KV pruning methods.
However, its accuracy remains lower than that of \SmartGen, underscoring the necessity of on-demand transfer.
We also observe an interesting phenomenon: the proactive-only baselines achieve higher accuracy than StreamingLLM on the Gemma-3-12B model, which uses a mix of sliding and full attention layers.
This is likely because Gemma-3-12B is more sensitive to important tokens in the full-attention layers, whose KV distributions do not align with StreamingLLM's pattern.
\section{Related Work}

\subsection{LLM Inference Systems}
% Orca ⇒ vLLM ⇒ chunked prefill、PD分离（参考distserve的第一段related work）
% The rapid advancement of LLMs has attracted increasing attention in terms of enhancing the efficiency of LLM inference systems in many areas, \eg 
\noindent
The rapid advancement of LLMs has attracted increasing attention on enhancing LLM inference systems in many areas, \eg 
request scheduling to satisfy SLO requirements~\cite{osdi2024sarathi, osdi2022orca, osdi25blitzscale, asplos2025helix, osdi24usher, sosp25prefillonly, osdi25nanoflow, osdi20deepak, sosp2025iccache, iclr2025preble, osdi2024fairness}, 
memory management to save GPU memory~\cite{sosp2023vllm, nips2024sglang, sosp25jenga, sosp2024powerinfer, eurosys25cacheblend, arxiv2026fusionrag, eurosys25pensieve, fast2026gpucheckpoint, sosp2025phoenixos}, 
resource disaggregation to attack the interference issue between the prefill and decoding computations~\cite{osdi2024distserve, fast2025mooncake, isca2024splitwise, icml2024dejavu, arxiv2024tetriinfer},
and hardware-software co-design that constructs effective kernels and accelerators to boost inference efficiency~\cite{isca2023olive, asplos25podattn, osdi25waferllm}.
\SmartGen focuses on optimizing the KV cache transfer when self-hosting disaggregated LLM inference systems.

\subsection{P/D Disaggregation}
\noindent
P/D disaggregation has become a widely adopted architecture for deployed LLM inference systems~\cite{fast2025mooncake, arxiv2024deepseekv3, nvidia-dynamo, sosp2023vllm, nips2024sglang}.
While this design enhances system scalability, it poses challenges in inter-node KV cache transfer~\cite{isca2024splitwise, osdi2024distserve, sigcomm25hack}.
DistServe~\cite{osdi2024distserve} introduces additional layer placement constraints to force KV cache transfer to occur only within a node, which limits the flexibility of resource disaggregation.
Splitwise~\cite{isca2024splitwise} and D{\'{e}}j{\`{a}}Vu~\cite{icml2024dejavu} propose to overlap the transfer with the prefill computation.
This could not fundamentally address the transfer issue as analyzed in Section~\ref{sec:analysis}.
Mooncake~\cite{fast2025mooncake} relies on high RDMA bandwidth (\eg 800 Gbps per machine) to enable efficient KV cache transfer across nodes.
\SmartGen focuses on optimizing KV cache transfer with limited network bandwidth, and could be incorporated into these systems.

\subsection{KV Cache Management}
\noindent
There is a line of research that explores saving GPU memory footprint through KV cache management, \eg 
virtualization~\cite{sosp2023vllm, asplos2025vattention}, 
offloading~\cite{icml2023flexgen, atc2024cachedattention, eurosys2025hcache, asplos2025aqua, hpca25instattn, atc2025weaver}, 
quantization~\cite{sigcomm2024cachegen, isca2025oaken, nips2024kvquant, sigcomm25hack}, and 
sparsity~\cite{osdi2024infinigen, nips2023h2o, icml2024quest, arxiv2024clusterkv, iclr2025omnikv, iclr26kvcomm, arxiv25retroinfer}.
Works that adopt sparsity relate most to \SmartGen.
\SmartGen benefits from KV cache sparsity algorithms and leverages them to address the KV cache transfer challenge in self-hosted disaggregated LLM inference.
Quantization-based methods like HACK~\cite{sigcomm25hack} are orthogonal to \SmartGen and can be applied in a complementary manner.
To our knowledge, \SmartGen is the \textit{first} work to enable seamless disaggregated LLM inference by exploiting the dynamic sparsity of KV caches.

\section{Conclusion}
\noindent
This paper identifies the stage-transition stall in disaggregated inference for self-hosted LLMs on the cloud.
We propose a selective KV cache transfer scheme, \SmartGen, that transfers essential KV cache entries across prefill and decoding stages to address this issue.
Experimental results verify the efficacy and efficiency of \SmartGen.

% \input{Sections/example_sections}

%-------------------------------------------------------------------------------
\bibliographystyle{plain}
\bibliography{main}

\end{document}